\documentclass[iop,revtex4]{emulateapj}

\usepackage{apjfonts}
\usepackage{lscape}

\usepackage[french, english]{babel}  

\usepackage{longtable}
\usepackage{amsmath}
\usepackage{natbib}
\usepackage{url}

\newcommand{\Msun}{\mbox{$M_{\odot}$}}
\newcommand{\Rsun}{\mbox{$R_{\odot}$}}
\newcommand{\Lsun}{\mbox{$L_{\odot}$}}

\newcommand{\Teff}{\mbox{$T_{\rm eff}$}}

\newcommand{\logg}{\mbox{$\log~g$}}
\newcommand{\Lbol}{\mbox{$L_{\rm bol}$}}

\newcommand{\apgt} {\ {\raise-.5ex\hbox{$\buildrel>\over\sim$}}\ }
\newcommand{\aplt} {\ {\raise-.5ex\hbox{$\buildrel<\over\sim$}}\ } 
\newcommand{\kms}{\hbox{km~s$^{-1}$}}
\newcommand{\vsini}{\hbox{$v~\sin~i$}}

\slugcomment{Accepted to ApJ}

\begin{document}

\title{BANYAN. IV. Fundamental parameters of low-mass star candidates in nearby young stellar kinematic groups - Isochronal Age determination using Magnetic evolutionary models}

\author{Lison Malo$^{1,2}$\footnote{Based on observations obtained at the Canada-France-Hawaii Telescope (CFHT) 
which is operated by the National Research Council of Canada, the Institut National des Sciences de 
l''Univers of the Centre National de la Recherche Scientique of France, and the University of Hawaii. }, 
Ren\'e Doyon$^1$, Gregory A. Feiden$^3$, Lo\"{i}c Albert$^1$,  David Lafreni\`ere$^1$, \'Etienne Artigau$^1$, Jonathan Gagn\'e$^1$ and Adric Riedel$^4$}

\affil{$^1$D\'epartement de physique and Observatoire du Mont-M\'egantic, Universit\'e de Montr\'eal, Montr\'eal, QC H3C 3J7, Canada}
\affil{$^2$Canada-France-Hawaii Telescope, 65-1238 Mamalahoa Hwy, Kamuela, HI 96743, USA}
\affil{$^3$Department of Physics and Astronomy, Uppsala University, Box 516, SE-751 20 Uppsala, Sweden}
\affil{$^4$Department of Astrophysics, American Museum of Natural History, Central Park West at 79th Street, New York, NY 10024, USA}

\email{malo@cfht.hawaii.edu, doyon@astro.umontreal.ca } 

\begin{abstract}
Based on high resolution optical spectra obtained with ESPaDOnS at CFHT, we determine fundamental parameters (\Teff, R, \Lbol, \logg\ and metallicity)
for 59 candidate members of nearby young kinematic groups.
The candidates were identified through the BANYAN Bayesian inference method of \citet{2013malo}, which takes into account the position, proper motion, magnitude, 
color, radial velocity and parallax (when available) to establish a membership probability.
The derived parameters are compared to Dartmouth Magnetic evolutionary models and to field stars with the goal to constrain the age of our candidates.  
We find that, in general, low-mass stars in our sample are more luminous and have inflated radii compared to older stars, a trend expected for pre-main sequence stars. 
The Dartmouth Magnetic evolutionary models show a good fit to observations of field K and M stars assuming a magnetic field strength of a few kG, as typically observed for cool stars. 
Using the low-mass members of $\beta$Pictoris moving group, we have re-examined the age inconsistency problem between Lithium Depletion age and isochronal age (Hertzspring-Russell diagram). 
We find that the inclusion of the magnetic field in evolutionary models increase the isochronal age estimates for the K5V-M5V stars. 
Using these models and field strengths, we derive an average isochronal age between 15 and 28 Myr and we confirm a clear Lithium Depletion Boundary 
from which an age of 26$\pm$3~Myr is derived, consistent with previous age estimates based on this method. 
\end{abstract}

\keywords{Galaxy: solar neighborhood ---
Methods: statistical --- 
Stars: distances, kinematics, low-mass, moving groups, pre-main sequence ---
Techniques: spectroscopic }

\section{Introduction}
 
In general, the determination of fundamental parameters, \Teff, R, \Lbol, \logg\ and metallicity, 
for a single star requires measurements of a trigonometric distance and an interferometric 
stellar diameter, as well as accurate photometric and spectroscopic observations.
Fundamental parameters have been derived for several old (field) low-mass stars as demonstrated by 
recent works \citep{2008casagrande, 2011casagrande, 2012boyajian, 2013rajpurohit, 2013mann}. 
However, little is known for the young population as there are relatively few young low-mass stars that 
have been unambiguously identified in the solar neighborhood, although the number of candidates 
is rapidly increasing \citep[e.g.][]{2014kraus,2014riedel, 2013malo, 2013rodriguez, 2012shkolnik, 2011rodriguez, 2008torres, 2004zuckerman}, and 
none have had their radii measured directly using interferometry.

Of all fundamental parameters, the age is probably the most difficult to constrain because its 
determination inevitably relies either on model-dependent methods (e.g., isochrone fitting, 
gyrochronology, the Li depletion boundary; LDB) or on kinematic traceback techniques for stars that are members of young 
co-moving groups \citep[][ and references therein]{2010soderblom, 2013soderblom}. 
In principle, age estimates from all these methods should be consistent but many studies have unveiled some inconsistencies. 
For example, LDB age is systematically greater than the isochronal age, a tend that is independent of the evolutionary model used.
This discrepancy perhaps suggests that other physical factors (e.g. metallicity, magnetic field strength, accretion history) are needed 
to fully account for the observational properties of young stars. 
Investigating the fundamental properties of young low-mass stars is strongly motivated by the fact that a significant \citep{2014barnes} 
fraction of nearby K and M dwarfs host exoplanets \citep{2007udry,2008casagrande}. 
The derived properties of those exoplanets rely on a good knowledge of the fundamental parameters of their host stars, namely their mass, 
effective temperature, radius, metallicity and, not least, their age.

This paper is part of a large program aimed at finding and characterizing low-mass stars in Young Moving Groups (YMG). 
In \citet[][hereafter Paper I]{2013malo}, we identified more than 150 highly probable members of young co-moving groups. 
We presented a Bayesian analysis, coined BANYAN, using kinematic and photometric information to infer the membership 
probability for a sample of low-mass stars showing strong H$\alpha$ and X-ray emissions. 
This analysis tool also provides a prediction for the most likely distance and the radial velocity of the candidates assuming that they are true members.  

In Malo et al. (accepted; hereafter Paper II), follow-up radial velocity (RV) observations were secured to show that a large fraction 
of the candidates have measured RVs matching the predictions, strenghtening the case that these are indeed genuine co-moving members. 
Several of them have now been confirmed as bona fide members based on recent parallax measurements \citep{2014riedel, 2012shkolnik}.
In paper II, we also showed that these young star candidates have unusually high X-ray luminosities and high rotational (\vsini)  
velocities compared to field counterparts. 
Thus, while a parallax will ultimately be needed to confirm their membership, 
those strong candidates already deserved further investigations. 

This paper is focused on the physical characterization of these stars with a strong emphasis to investigate how the magnetic fields 
affect the physical properties of low-mass stars. 
We use new high-resolution optical spectroscopy along with atmosphere models and various data from the literature to constrain the fundamental 
parameters of our young star candidates.  
The inferred physical properties are compared with Pre-Main Sequence (PMS) Dartmouth Magnetic evolutionary models\footnote{http://stellar.dartmouth.edu/models} 
from \citet{2012bfeiden} and \citet{2013feiden} with the goal of constraining the age and strengthening the case that those candidates 
are genuine members of their respective moving group. Our data are used to construct an Hertzsprung-Russell (HR) diagram of the $\beta$PMG extending 
further at the low-mass end, enabling an estimate of an isochronal age. 
New Li measurements are also used to derive an Lithium Depletion Boundary (LDB) age estimate.
We show that the age inconsistency problem can be partly solved, if the $\beta$PMG members have magnetic field strengths of 2.5~kGauss.

\section{Sample and Observation} \label{chap:deux}

A detailed description of our initial search sample was presented in Papers I and II. 
In summary, the sample includes low-mass stars (K5V-M5V) showing chromospheric X-ray and 
H$\alpha$ emissions, all with reliable $I_{c}$ photometry and proper motion measurements ($<$~0.2~mag and $>$~4~$\sigma$).
The sample comprises 920 stars, of which 75 were previously identified as young in the literature.
All candidates were considered for membership in the seven closest ($<$~100\,pc) and 
youngest ($<$~100\,Myr) comoving groups : TW Hydrae Association \citep[TWA;][]{1989delareza},  
$\beta$ Pictoris Moving Group \citep[$\beta$PMG;][]{2001bzuckerman}, 
Tucana-Horologium Association \citep[THA;][]{2000zuckermanwebb,2000torres},  
Columba Association \citep[COL;][]{2008torres}, 
Carina Association \citep[CAR;][]{2008torres},  Argus Association \citep[ARG;][]{2008torres} 
and AB Doradus Moving Group \citep[ABDMG;][]{2004zuckerman}. 

Applying our Bayesian analysis to this sample, 247 candidate members were found with 
a membership probability ($P$) over 90\%, amongh which 50 were already proposed as
candidate members in the literature.
In Paper II, the membership of 130 candidates was strengthened through radial and projected rotational
velocity measurements obtained via infrared and$/$or optical high-resolution spectroscopy.

\subsection{Definition of a bona fide member} \label{chap:deuxun}

As defined in this paper, a bona fide member is one that has all $XYZUVW$ parameters known from parallax, 
radial velocity and proper motion mesurements consistent with a high membership probability 
to a given YMG, as determined by various tools such as Bayesian inference \citep{2013malo, 2014gagne} and 
the convergent point analysis \citep{2014rodriguez,2013rodriguez}. 
A bona fide member is also required to display youth indicators. 
The most common indicator is the presence of Li, but this diagnostic is restricted to early M dwarfs younger than a few 10$^7$ yr since Li is rapidly 
depleted, especially in fully convective stars \citep{2001randich}.  
For older early M dwarfs, the location in the color-magnitude diagram is the only way to constrain their age. 
In paper II, we showed that bona fide low-mass members of YMGs show unusually high X-ray luminosities and rotational velocities 
which can be used as an independent youth indicator in the age range $\sim$10-100 Myr. 
Spectroscopic evidence of low-gravity \citep[e.g. NaI, KI;][]{2014riedel, 2003gorlova} is another useful youth indicator for mid-late M dwarfs. 

In summary, the youth of low-mass stars can be assessed through the following indicators: unusually high luminosity (bolometric, X-ray, and UV when available) compared 
to old stars of the same temperature (spectral type), unusually high rotational velocity, Li detection (depending on spectral type and age) and the gravity-sensitive NaI and KI lines. 
Because the interpretation of the observed luminosity is different in the case of an unresolved multiple system, 
RV monitoring and high-contrast imaging should be persued to 
identify binary systems within the proposed bona fide members. 

\subsection{New bona fide members} \label{chap:deuxdeux}

The RV measurements of Paper II combined with recent parallax measurements enable the identification of three new bona fide members.
The proposed three new bona fide members are all in the $\beta$PMG: J2033-2556 (M4.5V), J2010-2801 (M2.5+M3.5) and J2043-2433 (M3.7+M4.1). 
They all have a membership probability (P$_{v+\pi}$)  greater than 90\%, high X-ray luminosity typical of $\beta$PMG members
and they also show signs of low gravity \citep{2014riedel}.  
J2033-2556 has multi-epoch RV measurements ruling out a binary system with good confidence. 

\subsection{Ambiguous and Uncertain Members} \label{chap:deuxtrois}

We highlight six other candidates with known parallax whose membership is either ambiguous or uncertain for various reasons. 
Because of these raisons, we take the conservative approach of excluding these targets from the results presented in this paper.

{\bf 2MASSJ01351393-0712517} is a spectroscopic binary (M4.5V) identified in Paper II as a strong candidate member of Columba, but a recent parallax 
measurement \citep{2012shkolnik} yields a higher, but still ambiguous, membership probability (P$_{v+\pi} $) of 76\% in favor of $\beta$PMG. 
While this star remains a good young star candidate,  its membership is not firm enough to declare it a bona fide member of $\beta$PMG. 

{\bf 2MASSJ01365516-0647379 (M4V+L0V)} is a visual binary that satisfies all criteria to be a bona fide member of $\beta$PMG. 
However, the atmosphere models fitting analysis presented in Section~\ref{chap:troisun} yields an effective temperature of $\sim$3500~K 
that appears to be too high for an M4V ($\sim$3100~K) 
even though its bolometric luminosity ($\log$ \Lbol=-2.2 \Lsun) is consistent with a $\beta$PMG membership. 
At such a luminosity and temperature, this star should show some Li but it does not.  

{\bf 2MASSJ14252913-4113323} is a M2.5V spectroscopic binary with a strong membership probability in $\beta$PMG.
This binary was first proposed to be a potential member of TWA (with a marginal kinematic fit) by \citet{2014riedel} using the EW Lithium absorption and the NaI~8200 index. 
However, as discussed in Paper II, this object is relatively distant (67~pc) and youth indicators (NaI, Li) are consistent with a membership to 
an association significantly younger than $\beta$PMG, perhaps the Scorpius-Centaurus complex \citep{1999dezeeuw}. 
While this star is certainly young, its membership is doubtful enough to not declare it a bona fide member of $\beta$PMG. 

{\bf HIP 11152} is a M3Ve with a high membership probability to $\beta$PMG.
However, the derived \Teff\ = 3906$\pm$20~K \citep{2013pecaut} is too hot for a star of this spectral type; \citet{2013pecaut} derived a spectral type of M1V for this star. 
At this temperature, the star appears under-luminous for a membership in  $\beta$PMG. 
Furthermore, no Li is detected in this star which is incompatible with all other $\beta$PMG members of similar \Teff\ that show some Li. 
If this star is a M3Ve, it most likely has a \Teff\ of $\sim$3100~K close to the Li boundary transition. 

{\bf 2MASSJ00233468+2014282 and 2MASSJ23314492-0244395} are candidate members (K7V(sb2), M4.5V) with a membership probability (P$_{v}$) under 90\%. 
These stars are good young star candidates, but their membership remains to low to consider them into this analysis.  

\subsection{Observations and Data Reduction} \label{chap:deuxtrois}

Since 2010, 54 young star candidates were observed with ESPaDOnS, a visible-light echelle spectrograph
at CFHT\footnote{CFHT program: 11AC13, 11BC08, 11BC99, 12AC23, 12BC24, 13AC23, 13BC33}.
Observations were done using the "star+sky" mode combined with the "slow" and "normal" CCD
readout mode.  
The spectra have a resolving power R $\sim$ 68,000 and covers the 3700 to 10500 \AA\ 
spectral domain over 40 orders.
The observations were obtained with individual exposures of 60 to 1800 seconds depending on 
the target brightness, yielding typical signal-to-noise ratios (S$/$N) of $\sim$ 80-120 per pixel (2.6 \kms).

All observations were processed by CFHT using UPENA 1.0, an in-house software
that calls the Libre-ESpRIT pipeline \citep{1997donati}. 
We used the processed spectra with the unnormalized continuum and
the automatic wavelength solution inferred from telluric lines.  

Since no telluric standards were observed within the same night as the science targets, telluric
correction was achieved using 40 A0V spectroscopic standards secured from other ESPaDOnS programs
available through the CFHT data archive. 
The telluric correction involves the following steps.
First, we find the spectroscopic standard obtained through atmospheric conditions as close as
possible to that of the target star.
This choice is done by performing a linear combination of several telluric standards minimizing
the ratio of absorption depth for several prominent telluric lines.
Second, hydrogen absorption lines are removed from the telluric standard spectrum by fitting and 
dividing out a Voigt profile to each line. 
Finally, the target spectrum is divided by the corrected telluric standard spectrum, followed by the
division of a blackbody curve with the effective temperature of the chosen spectroscopic standard.

In order to compare ESPaDOnS spectra with atmosphere models, the spectra were flux-calibrated
by integrating target spectra with appropriate spectral response curves and scaling the 
results to match the apparent fluxes of the target star through various photometric bands. 
The apparent magnitudes came from various sources: SDSS-DR9 \citep{2011adelman},  
UCAC4-APASS \citep{2013zacharias}, Tycho \citep{2000hog}, DENIS \citep{1997epchtein}, 
\citet{2014riedel} or \citet{2002koen, 2010koen}.
The relative spectral response curves, effective wavelengths and zero points were taken from the Filter Profile 
Service\footnote{SVO: http://svo2.cab.inta-csic.es/theory/fps}, and
APASS filter transmission curves were kindly provided by Dr. Helmar Adler.

\section{Spectral analysis} \label{chap:trois}

\subsection{Fundamental Parameters Determination} \label{chap:troisun}

Fundamental stellar parameters, namely the effective temperature ($\Teff$), surface gravity ($\logg$),  
metalliticy ([M$/$H]) and stellar radius (R), were determined by fitting atmosphere models
to our spectra. 
The derivation of the stellar radius requires a distance estimate which, when a parallax is not available, is taken
to be the statistical distance (d$_{s}$) inferred from the method explains in Paper I.

We adopted the same spectral fitting analysis presented in \citet{2004amohanty} and 
\citet{2008mentuch} which consists of restricting the analysis to spectral regions  
strongly sensitive to surface gravity and effective temperature, namely the NaI and KI lines 
(see Table~\ref{tab:region}).

As stated in \citet{2011reyle} and \citet{2013mann}, the TiO opacity database is not complete for 
the BT-Settl models used here, hence TiO bands were not included in our analysis. 
Moreover, the NaI line at 589~nm is strongly affected by chromospheric emission lines, which may 
lead to a biased estimate of the effective temperature.

\begin{deluxetable}{lrrrrr}
\tabletypesize{\scriptsize}
\tablewidth{0pt}
\tablecolumns{6}
\tablecaption{Main properties of spectral region fitted \label{tab:region}}
\tablehead{
\colhead{Line} & \colhead{$\lambda$} & \colhead{$\Delta \lambda$} & \colhead{Temperature} & \colhead{Sensitivity} & \colhead{Refs.}\\
\colhead{element} & \colhead{(\AA)} & \colhead{(\AA)} & \colhead{(K)} & \colhead{} & \colhead{}
}
\startdata
Na I & 8183,8195  & 8150-8230 & 2500-3000 & logg, T & 1\\
K I  & 7665,7699  & 7660-7770 & 2500-3000 & logg, T & 2
\enddata
\tablerefs{(1) \citealp{2008mentuch}; (2) \citealp{2004amohanty} }
\end{deluxetable}

\subsection{Atmosphere Models and Fitting Method} \label{chap:troisdeux}

The candidate spectra were fitted with the BT-Settl atmosphere models \citep{2012allard} 
using solar abundances from \citet{2009asplund}.
These models are available for temperatures between 400 and 70,000~K, \logg\ between 
-0.5 to 5.5 and metallicity between -4.0 to +0.5.
For the purpose of our analysis, the atmosphere models considered were restricted to \Teff\ between 
2700 and 5000~K, \logg\ between 4.0 and 5.5 and [M$/$H] between -0.5 and +0.3, in steps of 100~K, 
0.5~dex, 0.3~dex, respectively. 

We have linearly interpolated atmosphere models separated by 100 K to construct a model grid with 50 K resolution
in \Teff; analogous averaging of models separated by 0.5 dex yields a grid with 0.25 dex resolution in \logg. 
The metallicity was interpolated at a value of -0.25 dex using the -0.5 and +0.0 atmosphere models.
This procedure was done to improve the numerical precision of the fitting procedure, as shown in \citet{2004bmohanty}

Prior to the fitting procedure, all target spectra were corrected for their heliocentric radial velocity as measured
in Paper II and the model spectra were convolved with a Gaussian kernel to match the resolving power and with a
rotational broadening profile to match the measured \vsini~ (see Paper II, and Table~\ref{tab:candprop}) of the targets.

The best model fit was determined by minimizing the goodness-of-fit parameter $G_{k}$ defined by \citet{2008cushing}
as :

\begin{alignat}{2}
G_{k} &= \sum_{i=1}^{n} W_{i} \left ( \frac{F_{obs,i}-C_{k}F_{k,i}}{\sigma_{obs,i}} \right )^{2}
\end{alignat}

where $F_{\rm obs,i}$ is the observed flux at wavelength $i$, $\sigma_{obs,i}$ is the associated uncertainty,
$F_{k,i}$ is the synthetic model flux for the same wavelength, and W$_{i}$ is the weight applied at each wavelength.

The parameter C$_{k}$ is set to minimize G$_{k}$ and corresponds to the value of $(R/d)^{2}$ where $R$ is the stellar radius and $d$ the distance to
the given star.
The subscript $k$ refers to a model with a given set of \Teff, \logg\ and [M$/$H].

Uncertainties on the derived fundamental parameters were determined as in \citet{2006casagrande,2008casagrande}
through Monte Carlo simulations; by repeating the above fitting procedure 51 times, each time with a different
random realization of the observed spectrum within the measurement errors (distance, flux calibration, gaussian noise on each spectral pixel). 
Typical uncertainties are 40~K, 0.15~dex, 0.05~$\Rsun$ for $\Teff$, \logg\ and $R$, respectively.

\begin{figure}[!hbt]
\epsscale{1.2}
\plotone{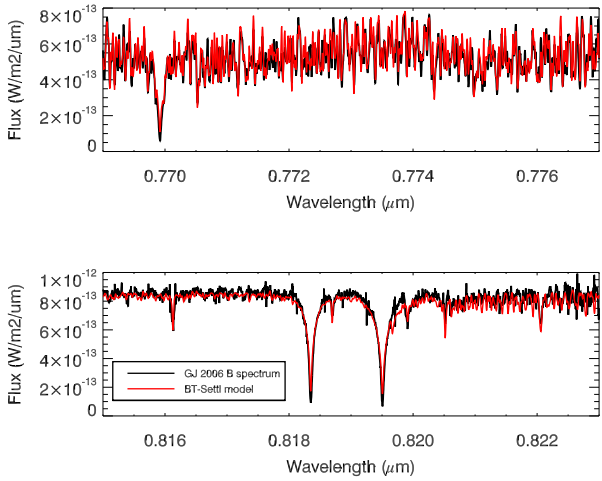}
\caption{\footnotesize{Example of fitting results for GJ 2006 B (2MASS~J00275035-3233238).} \label{fig:fitting_example}}
\end{figure}

The bolometric flux ($F_{bol}$) was estimated by integrating the best atmosphere model, 
which depends on $\Teff$, $\logg$ and metallicity, at the best radius found.
The bolometric luminosity given by $L_{bol} = 4~\pi~d^{2}~F_{bol}$ was derived from the statistical distance when 
parallax measurement was not available.
The inferred parameters for all candidate stars are given in Table~\ref{tab:candprop}.
Figure~\ref{fig:fitting_example} presents a comparison between observations and the best-fitting atmosphere model for the 
$\beta$PMG candidate member GJ~2006~B.

\subsection{Comparison to Stars of Known Parameters} \label{chap:troistrois}

For the purpose of validating our analysis, the same method was applied to five stars 
(GJ~880, GJ~205, HIP~67155 (GJ~526), GJ~687, GJ~411) with known parallax and fundamental parameters determined
independently by \citet{2012boyajian}, \citet{2013mann} and \citet{2013pecaut}.
The spectrum of HIP~67155 is from the CADC, but it was also obtained with ESPaDOnS and we applied the 
same analysis procedure described above.
As before, since no telluric standards were observed with the target observations, the same procedure
described in section~\ref{chap:deuxun} was used to find the best telluric standard.
As the ESPaDOnS CCD was replaced between semesters 2010B and 2011A, we selected telluric standards 
observed with the same CCD as the observations.

\begin{figure}[!hbt]
\epsscale{0.9}
\plotone{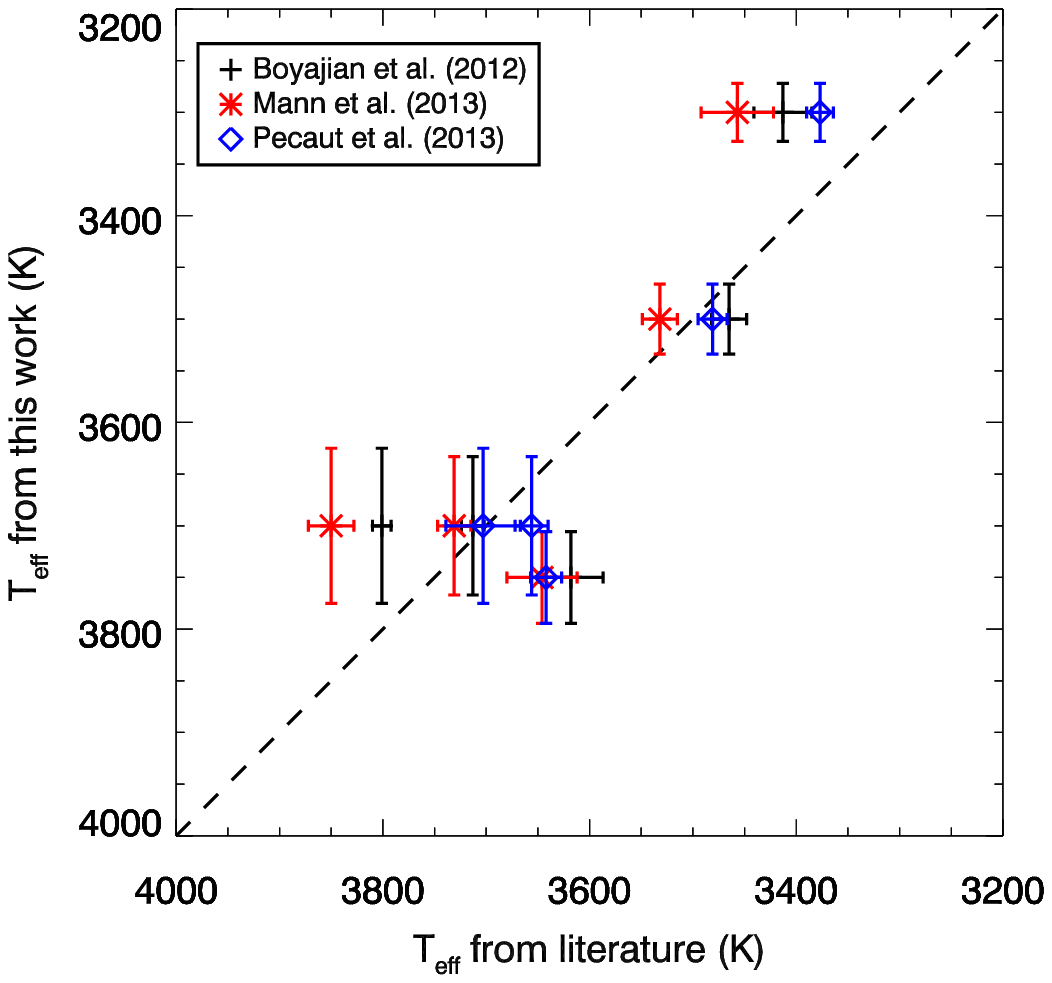}
\plotone{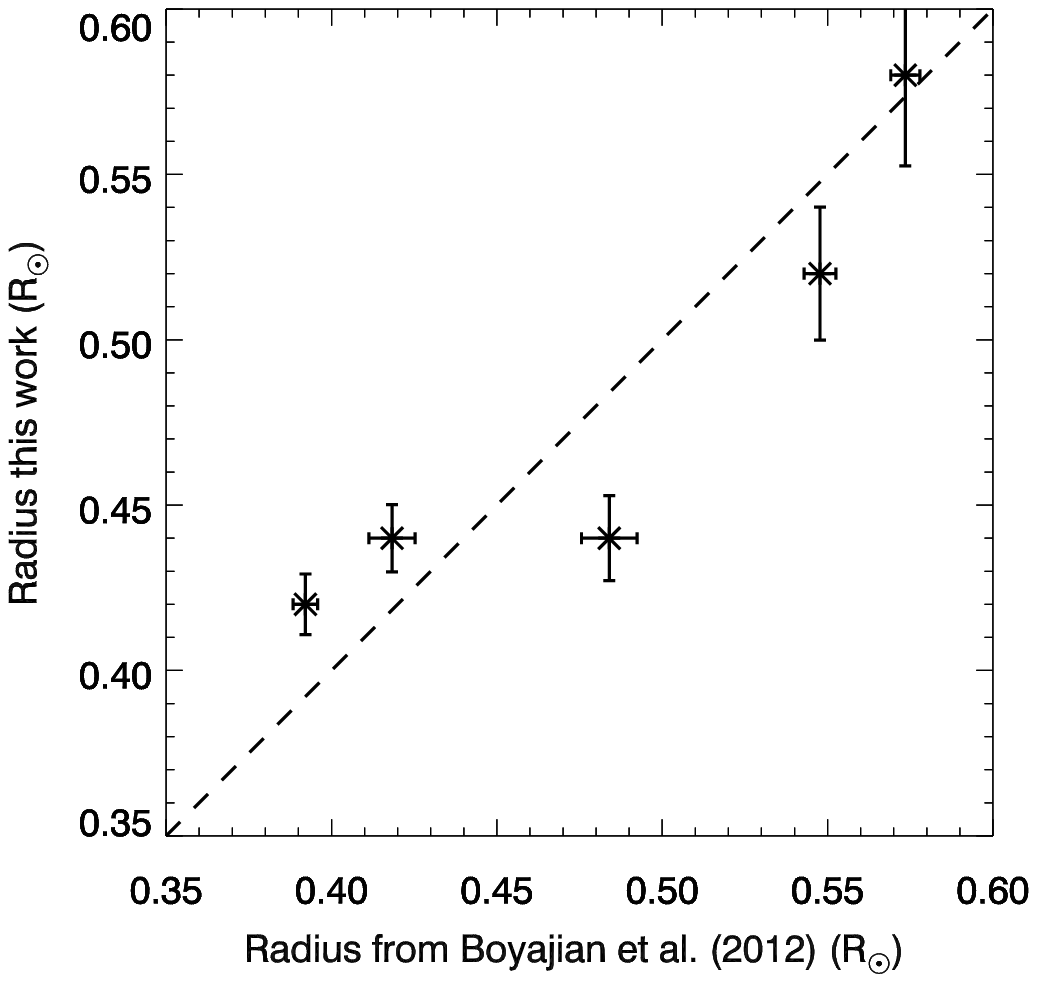}
\caption{\footnotesize{Top: Effective temperatures from this work as a function of \Teff\ from the literature.
Bottom: Estimated radii from this work as function of radii from the literature.} \label{fig:comparison_known}}
\end{figure}

Figure~\ref{fig:comparison_known} presents our estimated effective temperatures and radii compared 
to the literature measurements.
There is a good correlation between all estimates with a standard
deviation of 3\% and 5\% for \Teff\ and $R$, respectively. 

\subsection{Lithium Equivalent Width} \label{chap:troisquatre}

Our spectroscopic data includes the Li resonance line at  6707.8~\AA, a very important age indicator in young stars.  
The lithium equivalent width (EW) was measured using the following procedure. 
All spectra were first corrected for their respective heliocentric velocity. 

\begin{figure}[!hbt]
\epsscale{1.0}
\plotone{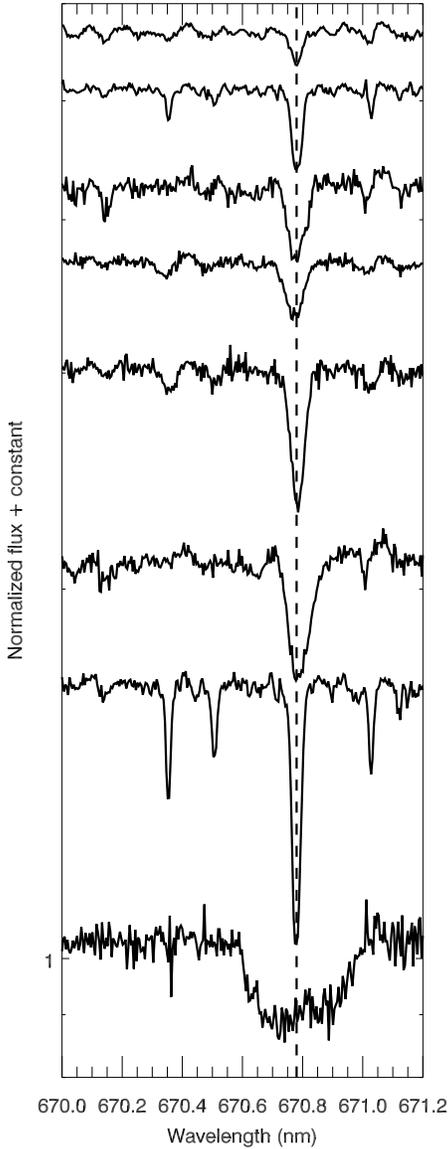}
\caption{\footnotesize{Example of Lithium line measurements for 
SCR1425-4113, HIP 11437, J2033-2556, J1923-4606, J0407-2918, J2110-2710, J0023+2014 and PMI04439+3723W (from bottom to top)} \label{fig:lithium_example}}
\end{figure}

Then the Li absorption feature was fitted between 6990 and 6710~\AA\ with a two-parameter function: 
the two parameters are the slope of the local continuum and the depth of the assumed Gaussian absorption feature.
The width of the assumed Gaussian absorption feature was set to the rotational velocity of the star (measured in paper~II) after convolution with the instrumental profile determined 
by fitting a Gaussian to a slow rotator featuring a high lithium EW.
Uncertainties were determined through a Monte Carlo analysis, i.e., by adding random Gaussian noise to the data and repeating the fitting procedure above. 
Figure~\ref{fig:lithium_example} presents examples of Li absorption lines spanning a wide range of rotational velocities and EW strengths. 
The resulting Li EW are given in Table~\ref{tab:candprop}. 
The reported uncertainties are statistical only and do not take into account possible systematic uncertainties associated with the location of the 
local continuum that may be biased by the faint $\sim$20 m\AA\,Fe line at 6707.4~\AA\, located $\sim$4 spectral resolution elements away from the lithium line. 
For this reason, we adopt a conservative 5$\sigma$ criterion for reporting upper limits.  

\section{Magnetic Evolutionary Models} \label{chap:cinqun}

We use the fundamental parameters inferred for our candidates and compare them with predictions from evolutionary models 
with the goal of confirming their youth and constraining their age. 
Comparisons are performed in the theoretical \Lbol--\Teff\ plane rather than a color-magnitude diagram in order to avoid
uncertainties related to color--\Teff\ transformations \citep{2012boyajian}.
Since low-mass stars generally show strong chromospheric and coronal emission associated with magnetic activity (c.f., Paper II), including effects due
to magnetic fields in the stellar models may be relevant. We therefore use the Dartmouth magnetic stellar evolution models \citep{2012bfeiden,2013feiden},
which are based on the models by \citet{2008dotter}. 
The Dartmouth magnetic stellar evolution code allows for the computation of both non-magnetic (i.e., standard) and magnetic stellar models, permitting comparisons.

Standard and magnetic models all have solar calibrated abundances, $Z = 0.0188$, $Y = 0.276$, and a solar calibrated mixing length parameter, $\alpha_{\rm MLT} = 1.884$.
The solar calibration differs slightly from that presented by \citet{2013feiden} because surface boundary conditions are now matched at an optical depth
of $\tau = 10$ for all masses. 
Magnetic perturbations are introduced using two formulations, one coined a rotational dynamo ($\alpha-\Omega$) approach and the other a 
turbulent dynamo ($\alpha^2$) approach \citep{2013feiden}. 
These two formulations do not refer specifically to actual solutions of the equations of magnetohydrodynamics, but were instead developed to capture relevant physical 
effects thought to be associated with each dynamo process. 
In particular, the rotational dynamo formulation considers the stabilizing influence a magnetic field  may have on thermal convection 
\citep[modified Schwarschild criterion;][]{1961Chandrasekhar}, while the 
turbulent dynamo probes the influence of a reduced convective efficiency by removing energy from convective flows \citep{2006chabrier}.

\begin{figure*}[!hbt]
\epsscale{1.0}
\plotone{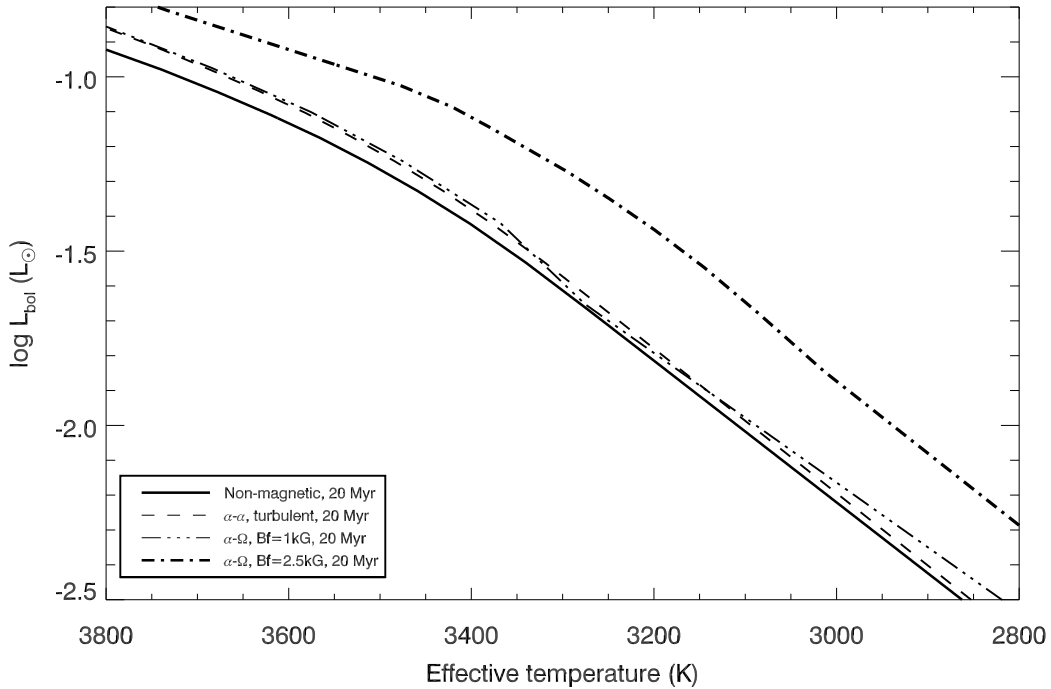}
\caption{\footnotesize{Bolometric luminosity difference between Rotational and Turbulent dynamos at an age of 20~Myr. 
The non-Magnetic Dartmouth model, turbulent dynamo, rotational dynamo with field strength of 1 and 2.5 kG are 
represented by the thick solid line, dashed line, long dashed line and dash dotted line, respectively.} \label{fig:model_diff}}
\end{figure*}

For investigations of main sequence eclipsing binaries, the effects of stabilization of convection and reductions of convective efficiency were
separated and associated with a rotational and turbulent dynamo, respectively, due to the magnitude of the magnetic field required to impart changes on
the structure of a star. 
Stabilizing convection requires interior magnetic fields of several hundred kilo-Gauss for early-M stars up to several mega-Gauss
for mid-M stars at or below the fully convective boundary \citep[][Feiden \& Chabrier, submitted]{2013feiden}. 
Magnetic field strengths of this magnitude cannot be generated purely by the conversion 
of kinetic energy from convection into magnetic energy. 
An input of energy from rotation would be required. 
However, removing kinetic energy from convection without explicitly modifying the Schwarzschild criterion can impart the same structural changes as stabilizing convection, but with
effects that correspond to magnetic field strengths in the range of several tens of kilo-Gauss \citep{2008browning, 2006chabrier}. 
Therefore, it was reasonable to separate the two effects and attribute them to separate ``dynamo mechanisms'' \citep{2013feiden}.

The two effects on stellar structure, however, need not be strictly independent.
For example, if a turbulent dynamo produces a magnetic field of a given magnitude drawn from kinetic energy in convection, that magnetic field can 
have a stabilizing effect on the convective flows. 
This is especially true for pre-main-sequence stars that may generate their magnetic field via a turbulent dynamo, but have physical conditions that make stabilizing 
convection just as relevant of a process when the magnetic field has a strength of only several tens of kilo-Gauss.

Given this, we use magnetic models calculated with both a turbulent dynamo formulation and the rotational dynamo formulation. 
Turbulent dynamo models have the radial magnetic field strength profile defined as a fraction, $\Lambda$, of the equipartition magnetic field strength at the given grid point within 
the model \citep[for details, see][Feiden \& Chaboyer 2014]{2013feiden}. 
Four values of $\Lambda$ were tested: $\Lambda = 0.5$, 0.75, 0.90, and 0.99, but only the last case will be discussed here.
At the masses considered in this study, equipartition of magnetic energy with kinetic energy in convection ($\Lambda = 1.0$) corresponds to a 
surface magnetic field strength of approximately 3.0~kG with interior fields around 50.0 kG. 
Rotational dynamo models were constructed with a dipole radial profile with a peak magnetic field occurring at a depth R = 0.5~\Rsun\ 
for fully convective stars and at the model tachocline, for stars with radiative cores. 
Three values of the surface magnetic field were used, $B_{\rm surf} = 1.0$, 2.0, and 2.5~kG, the latter being consistent with equipartition between magnetic and thermal pressures. 
Peak magnetic field strengths were on the order of 50~kG, and thus could be plausibly generated by a turbulent dynamo \citep{2008browning}.
The magnetic field strengths are representative of the values (typically between 1 and 4~kG) measured in K \& M dwarfs \citep{2012reiners}. 

All three models are compared in the HR diagram of Figure~\ref{fig:model_diff}. 
Non-magnetic, magnetic $\alpha^2$, and magnetic $\alpha-\Omega$ with 1~kG show modest differences. 
On the other hand, the magnetic $\alpha-\Omega$ model with a field strength of 2.5~kG shows a significant luminosity enhancement of 
$\sim$0.3~dex (at \Teff\ below 3400~K) compared to the non-magnetic case and/or other models. 

Recent theoretical studies \citep{2008browning, 2012gastine} and observational studies \citep{2006donati, 2008morin} have shown that 
fully convective objects with high stellar rotation should be able to produce large-scale magnetic field.
Therefore, the discussion above will maintly focus on comparing fundamental parameters of low-mass star candidates to the magnetic $\alpha-\Omega$ model.

\section{Results} \label{chap:cinq}

The analysis below includes the new data presented in this work as well as data from other sources. 
The properties of all stars are compiled in Table~\ref{tab:candprop}.  
The sample includes stars of various status: bona fide members, candidate members with parallax but ambiguous membership (see Section~\ref{chap:deuxdeux}), 
and candidates without parallax. 
Of all 59 stars studied spectroscopically through this work (54 candidates + 5 stars from E. Shkolnik 
observing program), 7 turned out to be field contaminants. 
The properties of those stars are given in Table~\ref{tab:fieldprop}. 
They will serve, along with other data from the literature, as an old sample for comparison. 

\subsection{Hertzsprung-Russell Diagram} \label{chap:cinqun}

\begin{figure*}[!hbt]
\epsscale{1.15}
\plotone{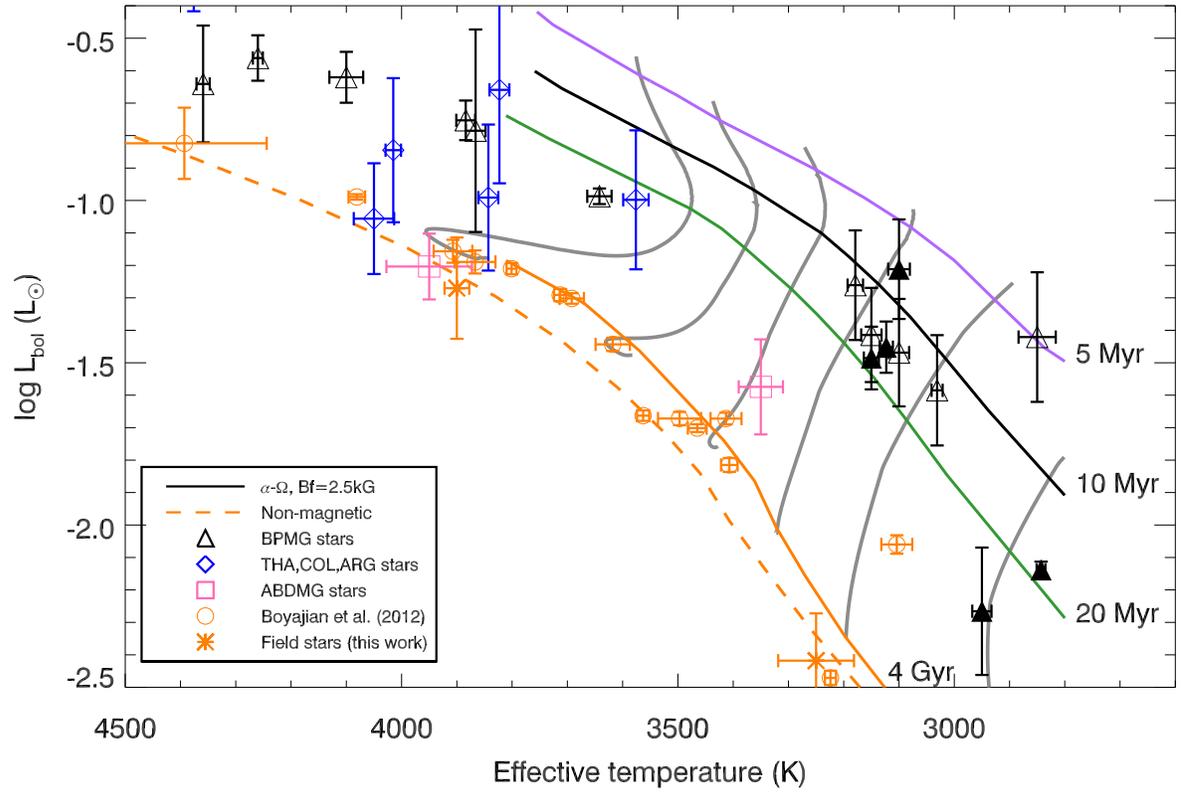}
\plotone{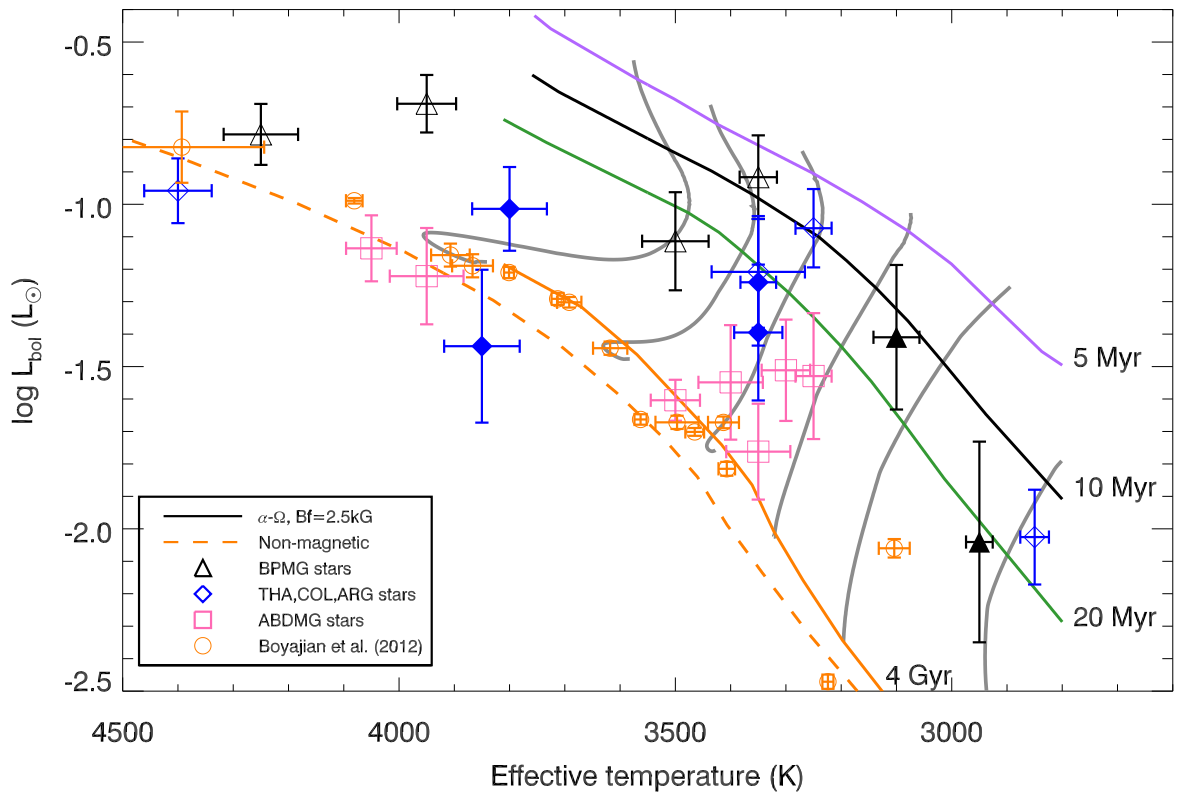}
\caption{\footnotesize{Top panel: Bolometric luminosity as a function of effective temperature for
young bona fide members. Binary stars are shown with filled symbols and their luminosity was reduced by a factor of two in this figure (see text). 
The non-magnetic and magnetic $\alpha-\Omega$ Dartmouth model with field strength 2.5~kG are represented by dashed and solid lines, respectively. 
The magnetic Dartmouth isochrones of 5 Myr, 10 Myr, 20 Myr and 4 Gyr are represented by purple, black, green and orange color lines, respectively.
From right to left, grey mass tracks are represented for stars with masses between 0.1 and 0.6 \Msun.
Bottom panel: Same figure for candidate members only.} \label{fig:luminosity_all}}
\end{figure*}

Figure~\ref{fig:luminosity_all} presents the Hertzsprung-Russell (HR) diagram 
for our sample of bona fide members (top panel) as well as young bona fide members from the literature whose
fundamental parameters were estimated by \citet{2013pecaut}. 
Only stars with parameters accurate to better than 5$\sigma$ are considered. 
The bottom panel is the same diagram but for the candidate members only.
Several of our candidates are multiple systems, including several visual binaries, the majority of them uncovered through our radial velocity survey (Paper II). 
The bolometric luminosities of theses systems, given in Table~\ref{tab:candprop}, include all components. 
For the purpose of comparing them with single stars, we made the simplified assumption that they are equal-luminosity systems, 
hence their luminosity was reduced by a factor of two in Figure~\ref{fig:luminosity_all}. 
This approximation is reasonable since the median flux ratio of the visual binaries uncovered in Paper II is 0.72. 

Figure~\ref{fig:luminosity_all} also includes a sample of old M dwarfs whose fundamental parameters were determined interferometrically by \citet{2012boyajian}. 
The sample is complemented by two field stars with measured parallax from \citet{2012shkolnik} and archival ESPaDOnS spectra to which our analysis could be applied for deriving their fundamental parameters.  
As shown in Figure~\ref{fig:luminosity_all}, the inferred bolometric luminosities and \Teff\ for the old stars we analyzed agree reasonably well with the old sequence. 
Thus, we can be confident that our analysis is viable for deriving fundamental parameters of our young star candidates.

Older stars in Figure~\ref{fig:luminosity_all} are compared with a 4-Gyr Dartmouth magnetic and non-magnetic evolutionary models. 
For the non-magnetic case, one can note a discrepancy of about 0.1~dex in $\log~\Lbol$ between observations and evolutionary models for \Teff\ $>$ than 3500~K. 
The disagreement is somewhat larger for lower \Teff\ and has been noted before by \citet{2012boyajian} using other models \citep[e.g.][]{1998baraffe}. 
Interestingly, a magnetic $\alpha-\Omega$ model with a field strength between 1 and 2 kG \citep[as typically observed;][]{2012reiners} provides a very good fit to observations. 
{\it One of main results of this work is that the inclusion of magnetic field in evolutionary models provides a better fit to observations for K5V to M5V stars.} 

The young star data of Figure~\ref{fig:luminosity_all} are also compared with young isochrones of 5, 10 and 20~Myr for magnetic models of 2.5 kG. 
Note that the 100 Myr isochrone is very close ($\sim$0.1 dex above at \Teff\ = 3400~K) to the 4 Gyr isochrone.  
In general, there is a trend in Figure~\ref{fig:luminosity_all} (top panel) where young bona fide members show higher luminosities compared to older (field) stars. 
The trend is also as expected within moving groups in the sense that, at a given \Teff, candidate members of $\beta$PMG (10-20 Myr), are more luminous 
($\sim$0.5~dex) compared to older ABDMG members ($\sim$70-120 Myr). 
Moving groups of intermediate ages (THA, COL; $\sim$20-40~Myr) are also over luminous compared to ABDMG but they share similar luminosities with $\beta$PMG members.  

We also note that, qualitatively, a single isochrone does not provide a good fit for all \Teff. 
This is particularly obvious for $\beta$PMG for which a $\sim$20 Myr-old isochrone appears to fit the data reasonably well for \Teff\ greater than $\sim$3500~K whereas an 
isochrone younger than 5 Myr appears to better fit the data at lower \Teff.  
We shall come back to this point in section~\ref{chap:sixun}. 
The same trend is also seen for late-type members of ABDMG that appear over-luminous for their expected age, and for 
the candidate members sample (bottom panel).

\subsection{Radius-Effective Temperature Relation} \label{chap:cinqdeux}

\begin{figure*}[!hbt]
\epsscale{1.15}
\plotone{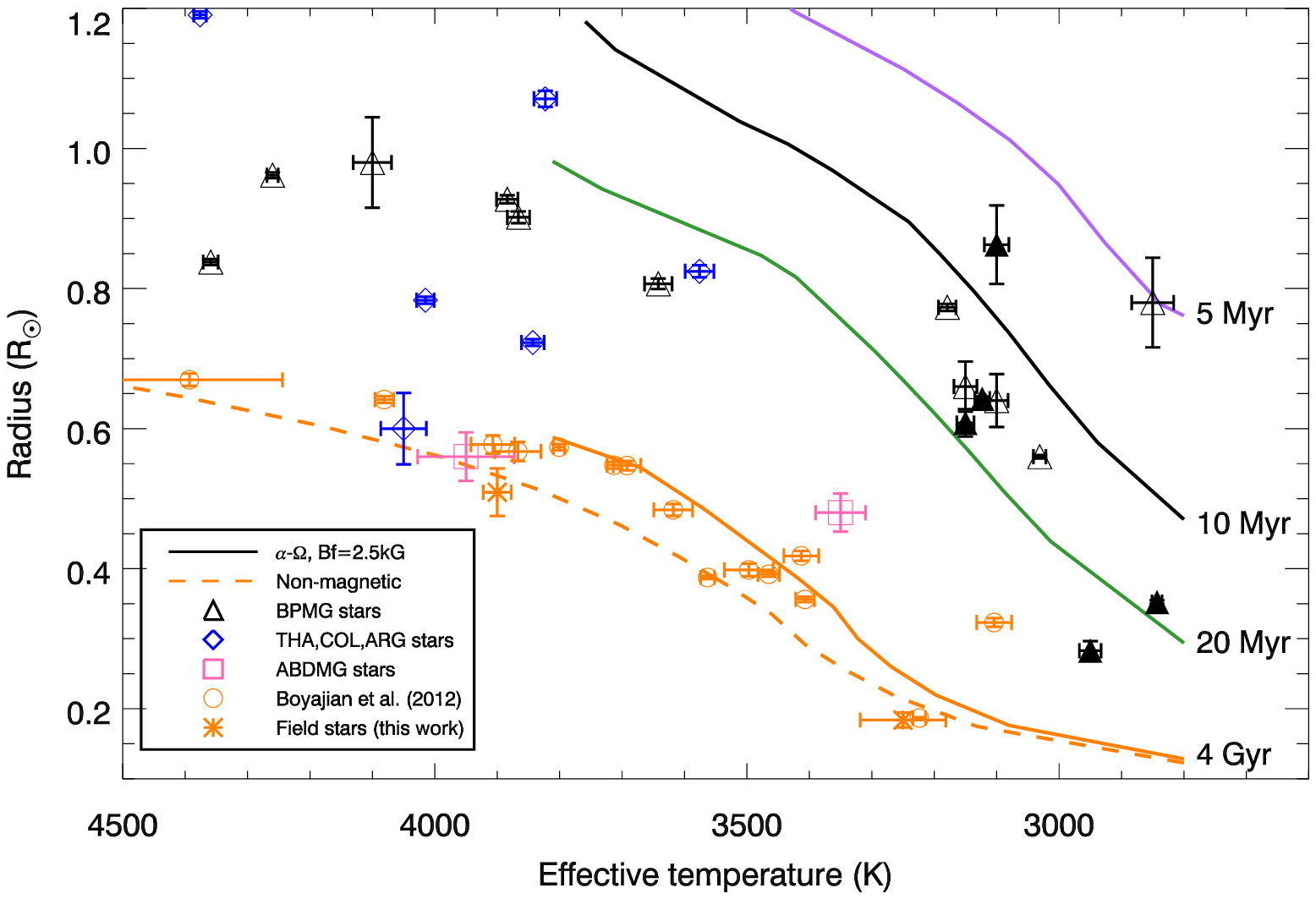}
\plotone{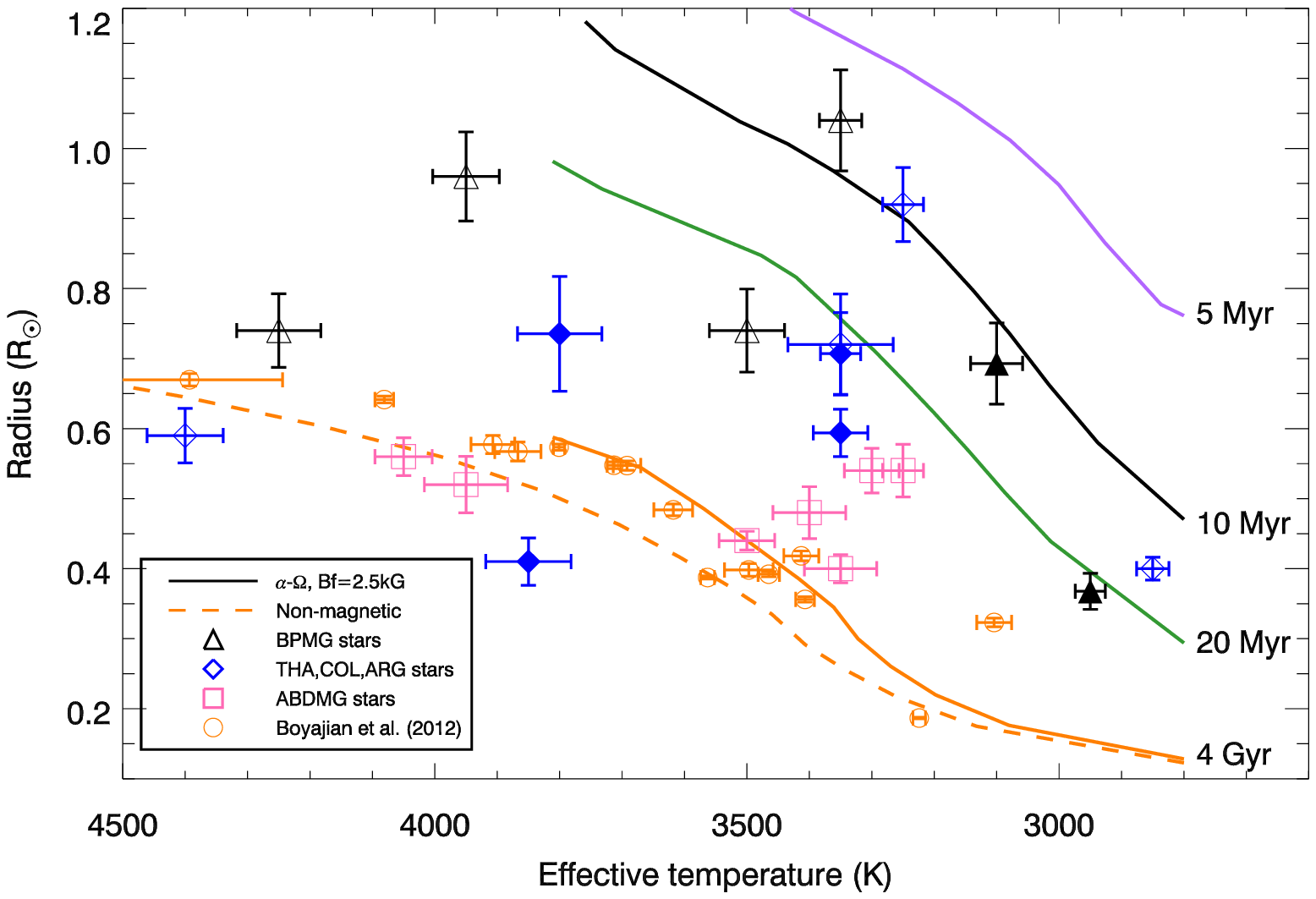}
\caption{\footnotesize{Top panel: Estimated radius as a function of effective temperature for
young bona fide members. Binary stars are shown with filled symbols.
The legend is the same as Figure~\ref{fig:luminosity_all}.
Bottom panel: Same figure for candidate members only.} \label{fig:radius_all}}
\end{figure*}

To complete this analysis, we present in Figure~\ref{fig:radius_all} the \Teff\ - Radius diagram for
the same samples used to construct the HR diagram.  
All young star candidates show inflated radii compared to Main Sequence (MS) stars, which is not surprising since the radius is effectively derived from the luminosity. 
The same trend observed in the HR diagram is also seen here, in that $\beta$PMG members have radii that are not consistent with a single isochrone. 
Late-type (\Teff\ $<$ 3500~K) stars have radii typically a factor of $\sim$2 larger than early-type members, which is consistent with the over 
luminosity factor of $\sim$4 (0.6 dex) observed in the HR diagram (see Figure~\ref{fig:luminosity_all}).

The main results of this analysis are the following : (1) as expected, bona fide members with known parallax show over luminosity, hence apparent inflated radii, compared to MS stars, and  
(2) no single theoretical isochrone can fit the observed fundamental parameters for low-mass members of $\beta$PMG.

\section{Discussion} \label{chap:six}

\subsection{The Age of the $\beta$Pictoris Moving Group} \label{chap:sixun}

$\beta$PMG is one of the youngest, nearest and best studied co-moving group in the solar neighborhood. 
The age of this group remains somewhat uncertain, although it is likely to be between 8 and 40~Myr. 
\citet{2013soderblom} presents a recent review of the various age estimates for $\beta$PMG along 
with the pros and cons of the various methods used so far.  
In summary, four different methods have been used: isochronal age, kinematic age, the Lithium Depletion Boundary (LDB) method, and the Li abundance. 

Using the HR diagram method,  \citet{1999barrado} and \citet{2001bzuckerman}  determined an isochronal age 
of 20$\pm$10~Myr and 12$^{+8}_{-4}$~Myr, respectively. 
The kinematic age method consists of tracing back the orbits of all members in time and finding when the volume of 
the group was smallest (i.e. when pairs of stars appeared to be closest to one another or to the group). 
Traceback ages of 11-12~Myr have been calculated by \citet{2002ortega} and \citet{2003song} 
while \citet{2007makarov} found a wider age range of 22$\pm$12~Myr using the flyby technique. 
Kinematic age methods have the distinct advantage of being independent of stellar models. 
On the other hand, they rely on the crucial  - and not necessarily correct - assumption that all members are coeval, and
that the group formed within a relatively small volume at birth. 
This method cannot be used reliably for groups (older than 50~Myr) as uncertainties in space velocities translate into large positional extent at birth. 
Furthermore, kinematic age methods suffer somewhat from some subjectivity in that the method does work only for some selected members. 
The third method, LDB, makes use of the fact that Li is rapidly depleted in low-mass stars and massive BDs, which in turn translates into a very 
sharp luminosity boundary between stars with and without Li. 
The LDB method \citep{1997bildsten,2006jeffries} is probably one of the most reliable methods for aging clusters as it relies on relatively simple physics 
and it is weakly dependent on the evolutionary models used \citep{2013soderblom}.  
The first LDB age estimate of 20~Myr for $\beta$PMG was reported by \citet{2002song}, based on the observation of the binary 
system HIP~112312 (M4V + M4.5V) whose primary shows a strong Li EW while the companion does not. 
\citet{2002song} argued that theoretical pre-MS evolutionary models were not able to simultaneously match the observed luminosity and the Li depletion 
pattern for both components and concluded that LDB ages are systematically overestimated. 
Recently, \citet{2013binks} determined an LDB age of 21$\pm$4~Myr based on several Li EW measurements.  
Finally, the Li abundance can also be used to derive the age through comparison with evolutionary and atmosphere models. 
\citet{2010macdonald} used this method to infer an age of $\sim$40~Myr for $\beta$PMG.

The results presented in this work give us the opportunity to derive both an isochronal age from low-mass 
stars and a new estimate of the LDB age thanks to new Li measurements. 
Morever, our high resolution spectroscopic observations provide useful constraints for identifying multiple systems that 
could affect the interpretation of the results. 

\begin{deluxetable}{lrl}
\tabletypesize{\scriptsize}
\tablewidth{0pt}
\tablecolumns{4}
\tablecaption{$\beta$PMG age from several indicators \label{tab:infobeta}}
\tablehead{
\colhead{Method} & \multicolumn{2}{c}{Derived sample age (Myr)} \\
\cline{2-3}
\colhead{} & \colhead{with $\pi$} & \colhead{all} 
}
\startdata
Isochrone T $>$ 3500 (no B field) & 15$\pm$1 & 21$\pm$3 \\
Isochrone T $<$ 3500 (no B field) & 5$\pm$1 & 5$\pm$1 \\
Isochrone T $>$ 3500 (1~kG) & 24$\pm$4 & 27$\pm$3 \\
Isochrone T $<$ 3500 (1~kG) & 6$\pm$1 & 6$\pm$1 \\
Isochrone T $<$ 3500 (2.5~kG) & 14$\pm$1 & 16$\pm$2 \\
LDB with binary & 30$\pm$2 & \ldots \\
LDB without binary & 26$\pm$3 & \ldots 
\enddata
\end{deluxetable}

\subsubsection{Isochronal Age from Low-Mass Stars} \label{chap:sixdeux}

Here we use the Darmouth evolutionary models to estimate the age of all stars with good fundamental parameters, i.e. the same sample of 
bona fide members used for constructing the HR diagram in Figure~\ref{fig:luminosity_all}. 
Figure~\ref{fig:hdr_age_bpmg} shows the estimated age for all stars as a function of \Teff, for non-magnetic and magnetic $\alpha-\Omega$ models (1 and 2.5 kG). 

Figure~\ref{fig:hdr_age_bpmg} provides a different way to illustrate that a single isochrone cannot match all of the observations. 
It also shows that the isochronal age depends on the magnetic field strength assumed. 
Table~\ref{tab:infobeta} gives a summary of various average age estimates for stars with effective temperature higher than or below 3500~K. 
First, we focus on the results using non-magnetic models.
All stars with \Teff\ higher than $\sim$3500~K show a similar and average age of 15$\pm$1~Myr while those with \Teff~$<$~3500~K are best 
fitted with an age of 4.5$\pm$0.5~Myr; those values were obtained by excluding all 5$\sigma$ outliers. 
Using the larger sample including candidates without parallax yields ages of 21$\pm$3~Myr for \Teff$>$3500 and 5$\pm$1~Myr for \Teff $<$ 3500~K. 
This discrepancy in age was also noted by \citet{2013binks} using Siess \citep{2000siess} evolutionary models (see their Figure~2) and 
\citet{2010yee} showed the same results using three different models (see their Figure~3).

Figure~\ref{fig:hdr_age_bpmg} shows that the age estimate does not vary monotonically with \Teff\ and instead shows an abrupt transition around 
$\sim$3500~K. 
This systematic age difference with \Teff\ (or mass) is probably not inherent to an age spread within the $\beta$PMG. 
While such an age dispersion is certainly possible, if not expected \citep[][ see also 6.1.2]{2013soderblom}, there is no good reason to expect very 
low-mass stars to be systematically younger than most massive ones.   
One can exclude ages as young as 4~Myr since all M dwarfs would be Li-rich at that age, which is not the case. 

If the age is not responsible for the observed excess luminosity in very low-mass stars, what could be the cause? 
One possibility is that those stars could be equal-luminosity unresolved binary systems that would be very difficult to detect spectroscopically. 
This binary hypothesis is far from satisfactory since it can account for only half of the observed luminosity excess and would imply that the binary frequency is higher at lower masses. 
It is unlikely that it could be the case. 

A more attractive alternative is to invoke the effect of magnetic field on evolutionary models, in particular the $\alpha-\Omega$ dynamo model for which the bolometric 
luminosity appears quite sensitive to magnetic field strength. 
As shown in Table~\ref{tab:infobeta} and Figure~\ref{fig:hdr_age_bpmg}, magnetic models tend to increase the inferred age. 
For stars with \Teff\ $<$ 3500~K, the age is increased from $\sim$5 to $\sim$15~Myr. 
This latter age is certainly much more probable than the former for $\beta$PMG. 
The apparent age difference between stars below and higher \Teff\ $\sim$3500~K could thus be explained if early-type stars have magnetic field strengths 
systematically lower compared to late-type ones. 
Using the compilation of magnetic field measurements of \citet{2012reiners}\footnote{http://solarphysics.livingreviews.org/Articles/lrsp-2012-1/}, 
one finds that old K0V-K5V stars have an average magnetic field of 
0.20$\pm$0.1~kG compared to 2.6$\pm$0.9~kG for M0V-M5V.  
Thus, there is a very significant trend for the magnetic strength to increase between K and M stars. 
However, this trend seems to disappear for young stars. 
Instead, using the same compilation, this time for PMS, one finds average magnetic field strengths of 2.2$\pm$0.7~kG and 2.4$\pm$0.8~kG for young K and M stars, respectively. 

The main conclusion from this discussion is that ignoring magnetic field systematically underestimates the ages derived from isochrones. 
The data presented here compared to the Darmouth magnetic evolutionary models suggests an isochronal age for $\beta$PMG likely between 15 and 28~Myr ($Bf$ =1~kG for T$>$3500~K). 

\subsubsection{An Age Gradient in $\beta$PMG ?} \label{chap:sixtrois}

The age derived above is an average value for the whole association and does not allow for a likely age spread. 
As discussed in \citet{2013soderblom}, the characteristic timescale $\tau_{SF}$ for the duration of a star formation event over a region of length scale $l$ should be:

$$\tau_{SF}\sim l_{pc}^{1/2}\,\,{\rm Myr}$$

Taking the current membership of $\beta$PMG and fitting an ellipsoid to the Galactic positions $XYZ$ of all members, one can estimate a characteristic radius of the group defined as $R_c\sim(abc)^{1/3}$, where $a$, $b$ and $c$ are the fitted semi-major axes of the ellipsoid. 
One finds $R_c\sim14$~pc or $\tau_{SF}\sim$5~Myr. 
Thus, from a theoretical point of view, $\beta$PMG members are expected to show an age spread of the order of $\sim$5~Myr. 

\begin{figure*}[!hbt]
\epsscale{1.2}
\plotone{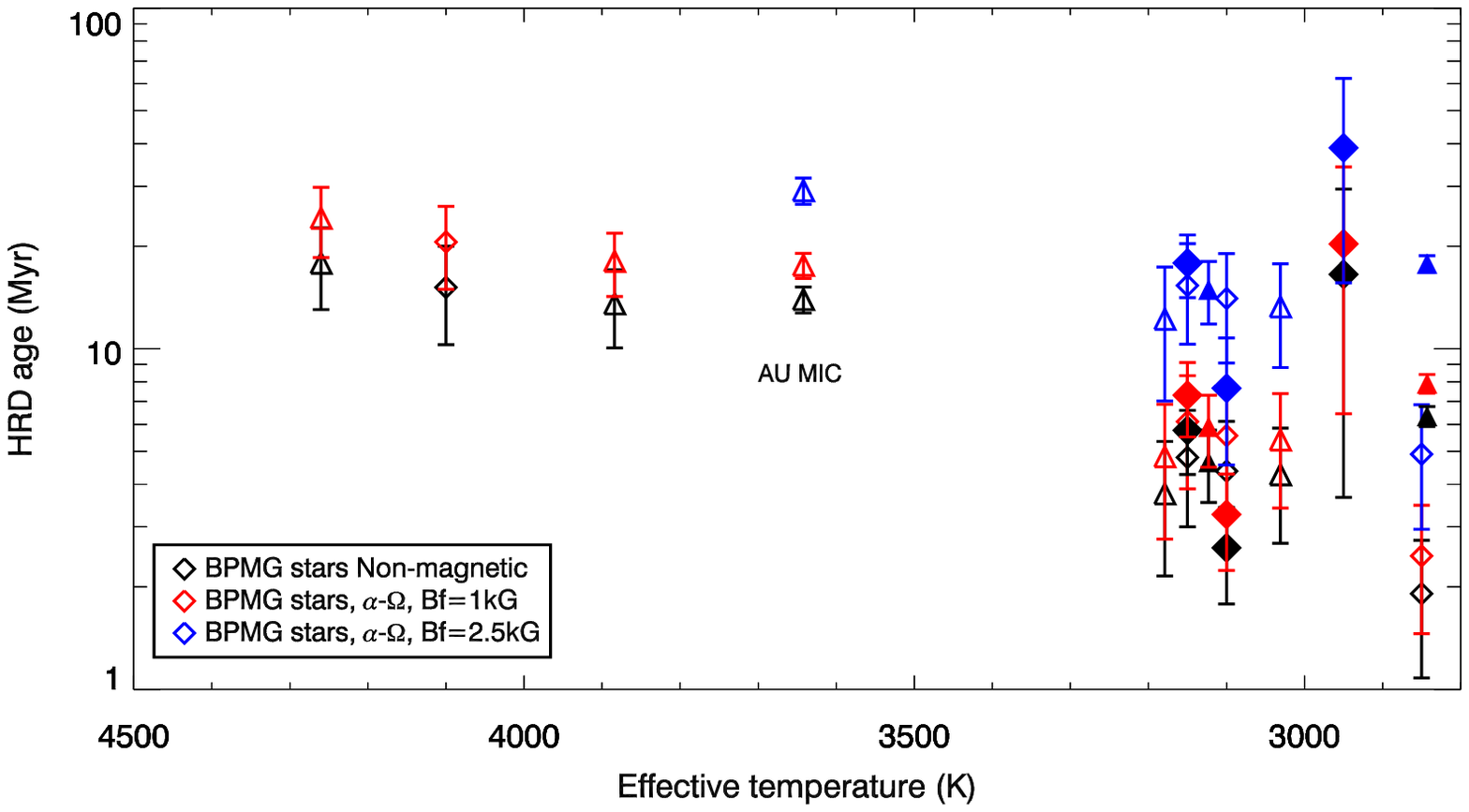}
\caption{\footnotesize{HR diagram age as a function of effective temperature for $\beta$PMG bona fide members.
Binary stars are shown with filled symbols.
Literature data from \citet{2013pecaut} is represented by triangle.
The star identified, AU MIC, has a measured magnetic field strength ($B~f$) of 2.3~kG; see text.} \label{fig:hdr_age_bpmg}}
\end{figure*}

There are two $\beta$PMG bona fide members for which magnetic field strengths are available; these measurements can be used for estimating the isochronal age of these individual stars based on magnetic evolutionary models. Those two stars are  
HIP~102409 (AU MIC), with a magnetic field strength $Bf$ of 2.3~kG \citep{1994saar}, and HIP~23200 (Gl~182), with $Bf$=2.5~kG \citep{2009creiners}. 
Only AU MIC is labeled in Figure~\ref{fig:hdr_age_bpmg}, since Gl~182 has a measured bolometric luminosity under our 5$\sigma$ criterion.
The magnetic $\alpha-\Omega$ model ($Bf$=2.5~kG) predicts an age of $\sim29\pm$3~Myr for AU MIC and $\sim26\pm10$ for GJ~182; the latter estimate requires a small extrapolation since our 
grid of magnetic $\alpha-\Omega$ models does not extend beyond \Teff=3800~K. 

Interestingly, both AU MIC and GJ~182 have a similar \Teff\ within $\sim$200~K and yet their Li EW differ by more than a factor of three. 
Indeed, as shown in Table~\ref{tab:candprop}, AU MIC ( \Teff=3642$\pm$22~K) has a relatively low Li EW (80~m\AA) compared to other $\beta$PMG members of similar effective temperature, for instance GJ~182 (270~m\AA; \Teff=3866$\pm$18~K) and HIP~23309 (360~m\AA; \Teff=3884$\pm$17~K). 
An age difference provides a natural explanation for explaining such a large difference in Li EW. 
This trend for AU MIC to have a relatively low Li-EW compared to other $\beta$PMG of similar \Teff\ may be an indication that AU MIC is somewhat older than a typical member of the $\beta$PMG.

\subsubsection{The Lithium Depletion Boundary Age} \label{chap:sixquatre}

\begin{figure}[!hbt]
\epsscale{1.2}
\plotone{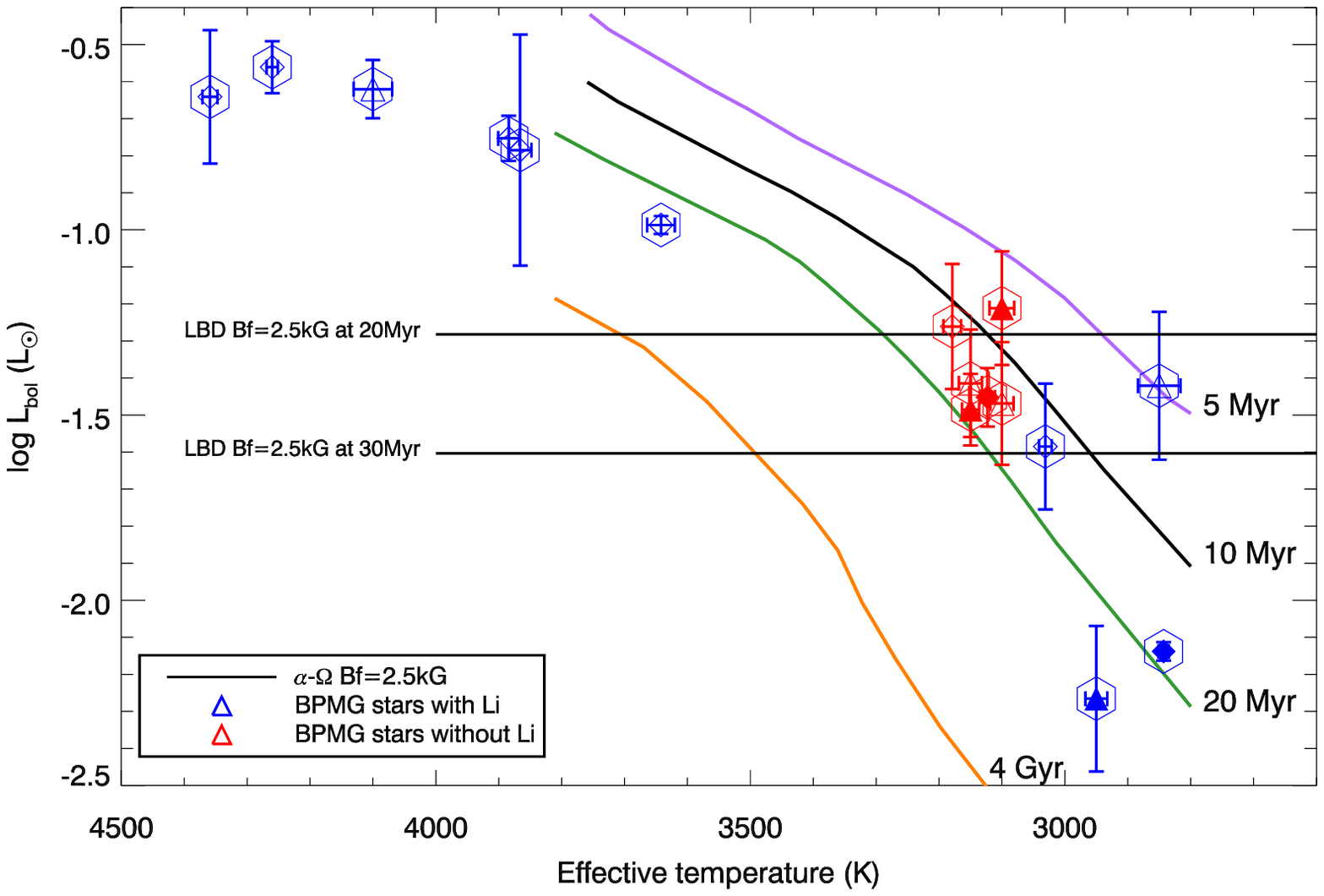}
\plotone{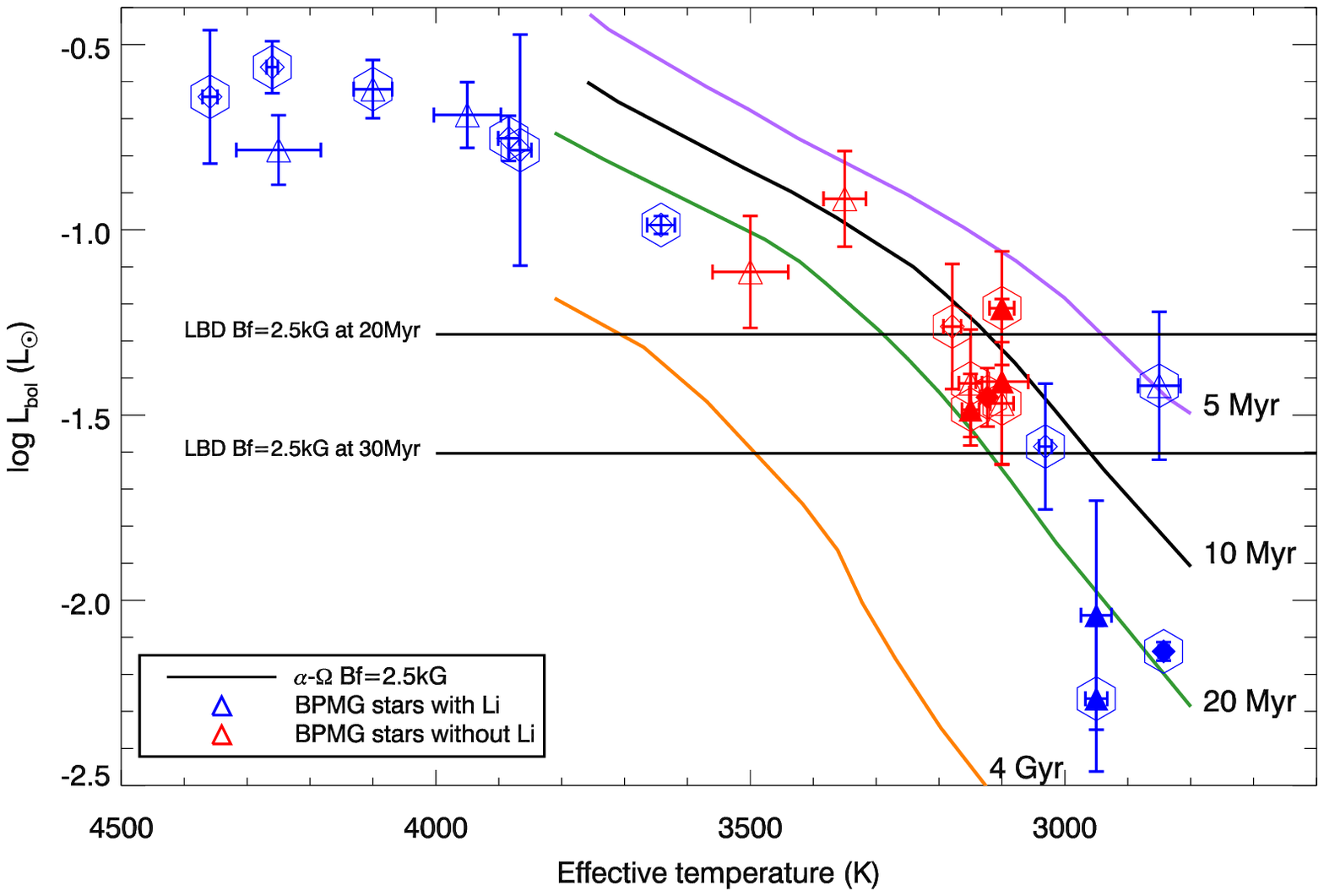}
\caption{\footnotesize{Top: Location of the Lithium Depletion Boundary (LDB) using luminosity as function of effective temperature for 
$\beta$PMG stars with known parallax measurements. Binary stars are identified with filled symbols.
Literature data from \citet{2013pecaut} is represented by diamonds.
Stars with and without lithium detection are represented by blue and red symbols, respectively.
The magnetic Dartmouth isochrones are as defined in Figure~\ref{fig:luminosity_all}.
Bottom: Same figure, but complemented with other strong $\beta$PMG candidates lacking parallax measurements.} \label{fig:lbd_bpmg}}
\end{figure}

Li is rapidly depleted in young low- and very low-mass stars, which in turn translates into a very 
sharp luminosity boundary between stars without and with lithium measurements.
Using a $K$-band color-magnitude diagram and Li EW measurements both from the literature and new observations, 
\citet{2013binks} determined an LDB age of 21$\pm$4~Myr for $\beta$PMG.  
Here, we repeat this analysis using a slightly different methodology. 
As before, we work in the \Lbol\ - \Teff\ plane and we use a maximum likelihood method for determining the LDB luminosity and its uncertainty. 
Figure~\ref{fig:lbd_bpmg} presents the HR diagram of all stars considered for the LDB analysis; this figure is similar to Figure~\ref{fig:luminosity_all}, except that stars are color 
coded to discriminate those with (blue symbols) and without lithium (red symbols). 
As usual, we treat two samples separately: one with measured parallaxes, and the other complemented with strong candidate members lacking parallaxes. 
All stars considered for the LDB analysis are identified in Table~\ref{tab:candprop}. 
The sample includes all data available from the literature including new observations presented here.
Figure~\ref{fig:lbd_bpmg} shows three kinds of objects: (1) K5V-K7V dwarfs with radiative core and convective envelope which still have high EW Li, 
(2) early-M dwarfs which may have radiative core or be fully convective object without lithium and (3) late-type fully convective M dwarfs with lithium detection. 

The bolometric luminosity $L_0$ at which the LDB occurs was determined using a maximum likelihood function. 
Let $L_i^a$ be the measured bolometric luminosity of the $i^{\rm th}$ target in the sample of $n$ stars with Li and 
$L_i^b$ be the measured bolometric luminosity for the $i^{\rm th}$ stars in the sample of $m$ stars without Li. 
The true bolometric luminosity of each star is characterized by a probability density function $f(l)$:
 
\begin{eqnarray}
f(l) = \frac{1}{\sigma_i\sqrt{2\pi}} e^{ -\frac{1}{2} \left ( \frac{l-l_i}{\sigma_i} \right )^2} \nonumber
\end{eqnarray} 

where $\sigma_i$ is the luminosity measurement uncertainty. 
Let $P^a$ be the probability that all stars with lithium have a true luminosity above $L_0$, 
$P^b$ the corresponding probability that stars without lithium have a true luminosity below $L_0$, and $\mathcal{L} (L_0)$  the probability density function for $L_0$, the LDB luminosity. 
We have,

\begin{eqnarray}
\mathcal{L} (L_0)=P^aP^b\nonumber
\end{eqnarray}

with

\begin{eqnarray}
P^a (L>L_0) = \prod_{i}^{n}\int_{L_0} ^{\infty} f(l_i^a)~dl \nonumber \\
P^b (L<L_0) = \prod_{i}^{m}\int_{-\infty} ^{L_0} f(l_i^b)~dl \nonumber \\
\end{eqnarray}

\begin{figure}[!h]
\epsscale{1.1}
\plotone{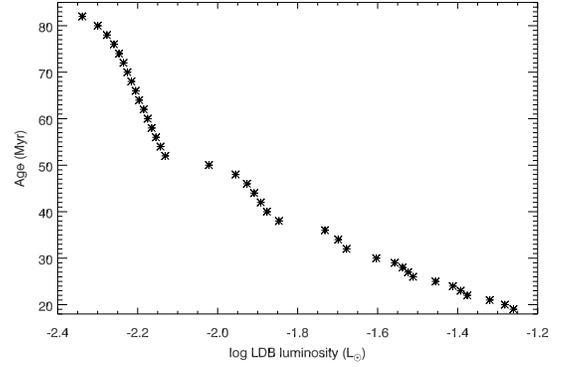}
\caption{\footnotesize{Age as function of the LDB Luminosity from Dartmouth Magnetic model (Bf = 2.5~kG) predictions.
The LDB luminosity is defined as the luminosity for which 99\% of the initial lithium has been depleted at that age.} \label{fig:lbd_age_bpmg}}
\end{figure}

Applying this methodology yields the most probable value and uncertainty for $L_0$, which can then be compared with 
theoretical predictions from the Darmouth models (see Figure~\ref{fig:lbd_age_bpmg}) to derive the corresponding LDB age. 
Here we adopt the same definition as \citet{2013binks} for $L_0$,  the luminosity at which 99\% of the initial Li abundance is depleted. 
Using the whole sample with parallax (11 stars) yields $L_0$=-1.59$\pm$0.06 \Lsun\ for a corresponding age of 30$\pm$2~Myr. 
The LDB age is potentially sensitive to the luminosity correction applied to binary systems, especially those that happen to be close to $L_0$. 
Excluding binary systems from the sample (5 stars left) yields $L_0$=-1.49$\pm$0.08 \Lsun\ for an age of 26$\pm$3~Myr.  
We adopt this value as the best LDB estimate since it is less likely affected by uncertainties associated with binarity.

Our LDB age of 26$\pm$3~Myr is consistent with the value of 21$\pm$4~Myr derived by \citet{2013binks}  using a different methodology and different evolutionary models. 
This is another illustration that the LDB age is relatively insensitive to the choice of evolutionary models, magnetic or not. 
It is encouraging that LDB age estimates for $\beta$PMG are in good agreement with the isochronal age range between 15 and 28~Myr. 

One should caution that the LDB age is not without systematic uncertainties associated with ill-understood Li depletion processes like rotation \citep{2008bouvier, 2009dasilva}, 
magnetic field \citep{2007chabrier} and early accretion history \citep{2009baraffe}. 
It has been suggested that strong differential rotation at the base of the convective envelope may be responsible for enhanced Li depletion in solar-type slow rotators \citep{2008bouvier}. 
Finally, accretion activity could potentially enhance lithium depletion rate 
during the first few Myr of star formation, in which case the LDB age should be regarded as an upper limit. 
Both the LDB method and magnetic evolutionary models yield a consistent age for $\beta$PMG of 26$\pm$3~Myr.

\subsection{The Age of Columba and THA} \label{chap:sixtrois}

We can apply the same analysis to other groups (Columba, THA and Argus) albeit with less accuracy, since there are fewer candidates with the required measurements 
than in $\beta$PMG (see Table~\ref{tab:candprop}). 
Since the data is sparse, we combined all three groups together assuming that they have approximately the same age, which is probably not inaccurate given they all 
share age estimates between 20 and 40 Myr.  
The sample comprises 12 stars, of which 6 have parallaxe measurements. 
As for $\beta$PMG, we divide the sample between early (T~$>$~3500~K) and late-type (T~$<$~3500~K) stars. 
The best isochrone fits yield 21$\pm$4~Myr for (T~$>$~3500~K) and 10$\pm$3~Myr for (T~$<$~3500~K). 
Again, the same trend is observed for late-type stars having relatively young ages. 
The age inferred from early-type stars is consistent with age estimates of these groups in the literature. 

\begin{figure}[!hbt]
\epsscale{1.2}
\plotone{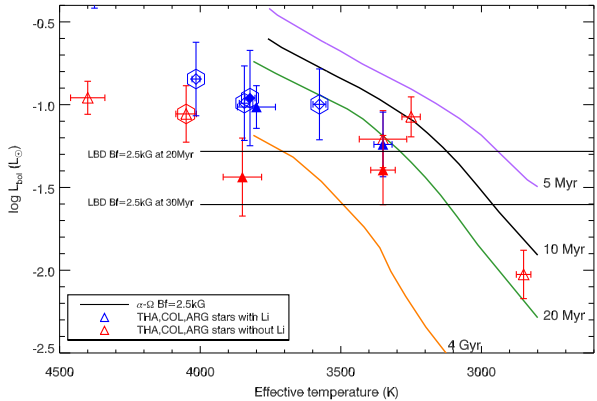}
\caption{\footnotesize{Locating the Lithium Depletion Boundary using Luminosity as function of effective temperature for 
THA, COL, ARG stars with known parallax measurements and candidates lacking parallaxes. Binary stars are identified with filled symbols.
Literature data from \citet{2013pecaut} is represented by diamonds.
The magnetic Dartmouth isochrones are as defined in Figure~\ref{fig:luminosity_all}.
} \label{fig:lbd_tha}}
\end{figure}

Since no late-type members with Li was found for these groups, only lower limits can be set for the average LDB age. 
Figure~\ref{fig:lbd_tha} shows the HR diagrams of the group. 
The significant number of late-type stars with undetected lithium spanning a wide range of luminosity provides a useful lower limit for $L_0$. 
Using the same maximum likelihood method described above, one can define a 3$\sigma$ lower limit for $L_0$ at which there is a probability of 
finding 99.7\% of stars above $L_0$. 
This limit is $L_0<$-2.29~\Lsun\ corresponding to an age lower limit of $\sim$ 79~Myr.

If we take into account only candidates of THA and COL, excluding the ARG member, the 3$\sigma$ lower limit 
for $L_0$ is -2.04~\Lsun\ corresponding to an lower limit age of $\sim$ 50~Myr. 
These lower limits are consistent with the fact that $\beta$PMG is very likely younger than those associations.

\section{Summary \& Concluding Remarks } \label{chap:sept}

We used multi-band optical photometry, high-resolution optical spectroscopy combined with
atmosphere model fitting to determine the fundamental parameters (\Teff, R, \Lbol, \logg\ and metallicity) of 59 candidates 
and bona fide members, to nearby young moving groups. 
In general, the candidates have higher bolometric luminosities and inflated radii 
compared to field old dwarfs.

We have explored the effects of the magnetic field on the age determination using 
the isochrone fitting method and the LDB method.
Using Dartmouth magnetic evolutionary models, we have shown that there is a good agreement between the 
models and the fundamental properties of old field stars by assuming a 
magnetic field strength of 2 kG, as typically observed or an old low-mass stars.
For $\beta$PMG members, isochronal ages inferred from magnetic models are 
systematically higher than those inferred from models that ignore the effects of 
magnetic field. 
We infer an isochronal age between 15 and 28 Myr using the magnetic models. 
This age pertains to the average age of the group.  
This relatively large age interval may reflects a dispersion in the magnetic properties
of the stars and/or a possible age spread within the association.
The LDB method yields an age of 26$\pm$3~Myr, consistent with previous estimates and
in agreement with the isochronal age derived in this work. 

The sample of young low-mass stars discussed in this work represents a relatively
 small fraction of all candidate members to nearby moving groups 
identified through Bayesian inference. 
Many other candidates have yet to be characterized and the vast majority remains
 to be identified. 
Bayesian inference has proven to be very efficient at identifying young low-mass stars
 and opens the exciting prospect of 
probing the sub-stellar and planetary mass regime of these groups. 

A paradigm shift for the study of young co-moving groups is expected when GAIA
 releases accurate astrometry (proper motion, parallax) and 
radial velocities for nearly all young low-mass stars 
in the solar neighborhood, enabling detailed characterization of these groups. 
While GAIA promises to revolutionize our understanding of nearby co-coming groups, other 
key measurements are needed to provide better observational constraints to evolutionary models, 
notably: interferometric radius measurements, high-resolution optical spectroscopy for atmospheric characterization, 
better radial velocity measurements (0.1 \kms) and magnetic field measurements through high-resolution infrared 
spectro-polarimetry, which will be possible with the SPIRou instrument under development \citep{2013delfosse}. 

\acknowledgments

We owe special thanks to Nadine Manset and the CFHT staff for helping to find archival telluric 
standards and doing the ESPaDOnS data reduction.  
Part of this work was possible thanks to Evgenya Shkolnik and Mike Liu who kindly provided several spectroscopic standards needed for our analysis. 
We also thank Tabetha Boyajian and Julien Morin for interesting discussions and advices.
Finally, we thank our referee for several comments which improved the quality of this
paper.

This work was supported in part through grants from the Fond de Recherche Qu\'eb\'ecois - Nature et Technologie and
the Natural Science and Engineering Research Council of Canada. 
This research has made use of the SIMBAD database, operated at Centre de Donn\'ees astronomiques de Strasbourg (CDS), 
Strasbourg, France. 
This research has made use of the VizieR catalogue access tool, CDS, Strasbourg, France \citep{2000ochsenbein}.

\bibliographystyle{apj}
\bibliography{references4}

\newcommand{\SortNoop}[1]{}
\begin{thebibliography}{87}
\expandafter\ifx\csname natexlab\endcsname\relax\def\natexlab#1{#1}\fi

\bibitem[{{Adelman-McCarthy} \& {et al.}(2011)}]{2011adelman}
{Adelman-McCarthy}, J.~K., \& {et al.} 2011, VizieR Online Data Catalog, 2306,
  0

\bibitem[{{Allard} {et~al.}(2012){Allard}, {Homeier}, \&
  {Freytag}}]{2012allard}
{Allard}, F., {Homeier}, D., \& {Freytag}, B. 2012, Royal Society of London
  Philosophical Transactions Series A, 370, 2765

\bibitem[{{Asplund} {et~al.}(2009){Asplund}, {Grevesse}, {Sauval}, \&
  {Scott}}]{2009asplund}
{Asplund}, M., {Grevesse}, N., {Sauval}, A.~J., \& {Scott}, P. 2009, \araa, 47,
  481

\bibitem[{{Bailey} {et~al.}(2012){Bailey}, {White}, {Blake}, {Charbonneau},
  {Barman}, {Tanner}, \& {Torres}}]{2012bailey}
{Bailey}, III, J.~I., {White}, R.~J., {Blake}, C.~H., {Charbonneau}, D.,
  {Barman}, T.~S., {Tanner}, A.~M., \& {Torres}, G. 2012, \apj, 749, 16

\bibitem[{{Baraffe} {et~al.}(1998){Baraffe}, {Chabrier}, {Allard}, \&
  {Hauschildt}}]{1998baraffe}
{Baraffe}, I., {Chabrier}, G., {Allard}, F., \& {Hauschildt}, P.~H. 1998, \aap,
  337, 403

\bibitem[{{Baraffe} {et~al.}(2009){Baraffe}, {Chabrier}, \&
  {Gallardo}}]{2009baraffe}
{Baraffe}, I., {Chabrier}, G., \& {Gallardo}, J. 2009, \apjl, 702, L27

\bibitem[{{Barnes} {et~al.}(2014){Barnes}, {Jenkins}, {Jones}, {Jeffers},
  {Rojo}, {Arriagada}, {Jord{\'a}n}, {Minniti}, {Tuomi}, {Pinfield}, \&
  {Anglada-Escud{\'e}}}]{2014barnes}
{Barnes}, J.~R., {et~al.} 2014, \mnras

\bibitem[{{Barrado y Navascu{\'e}s} {et~al.}(1999){Barrado y Navascu{\'e}s},
  {Stauffer}, {Song}, \& {Caillault}}]{1999barrado}
{Barrado y Navascu{\'e}s}, D., {Stauffer}, J.~R., {Song}, I., \& {Caillault},
  J. 1999, \apjl, 520, L123

\bibitem[{{Bildsten} {et~al.}(1997){Bildsten}, {Brown}, {Matzner}, \&
  {Ushomirsky}}]{1997bildsten}
{Bildsten}, L., {Brown}, E.~F., {Matzner}, C.~D., \& {Ushomirsky}, G. 1997,
  \apj, 482, 442

\bibitem[{{Binks} \& {Jeffries}(2013)}]{2013binks}
{Binks}, A.~S., \& {Jeffries}, R.~D. 2013, \mnras

\bibitem[{{Bobylev} {et~al.}(2007){Bobylev}, {Goncharov}, \&
  {Bajkova}}]{2006bobylev}
{Bobylev}, V.~V., {Goncharov}, G.~A., \& {Bajkova}, A.~T. 2007, VizieR Online
  Data Catalog, 908, 30821

\bibitem[{{Bouvier}(2008)}]{2008bouvier}
{Bouvier}, J. 2008, \aap, 489, L53

\bibitem[{{Boyajian} {et~al.}(2012){Boyajian}, {von Braun}, {van Belle},
  {McAlister}, {ten Brummelaar}, {Kane}, {Muirhead}, {Jones}, {White},
  {Schaefer}, {Ciardi}, {Henry}, {L{\'o}pez-Morales}, {Ridgway}, {Gies}, {Jao},
  {Rojas-Ayala}, {Parks}, {Sturmann}, {Sturmann}, {Turner}, {Farrington},
  {Goldfinger}, \& {Berger}}]{2012boyajian}
{Boyajian}, T.~S., {et~al.} 2012, \apj, 757, 112

\bibitem[{{Browning}(2008)}]{2008browning}
{Browning}, M.~K. 2008, \apj, 676, 1262

\bibitem[{{Casagrande} {et~al.}(2008){Casagrande}, {Flynn}, \&
  {Bessell}}]{2008casagrande}
{Casagrande}, L., {Flynn}, C., \& {Bessell}, M. 2008, \mnras, 389, 585

\bibitem[{{Casagrande} {et~al.}(2006){Casagrande}, {Portinari}, \&
  {Flynn}}]{2006casagrande}
{Casagrande}, L., {Portinari}, L., \& {Flynn}, C. 2006, \mnras, 373, 13

\bibitem[{{Casagrande} {et~al.}(2011){Casagrande}, {Sch{\"o}nrich}, {Asplund},
  {Cassisi}, {Ram{\'{\i}}rez}, {Mel{\'e}ndez}, {Bensby}, \&
  {Feltzing}}]{2011casagrande}
{Casagrande}, L., {Sch{\"o}nrich}, R., {Asplund}, M., {Cassisi}, S.,
  {Ram{\'{\i}}rez}, I., {Mel{\'e}ndez}, J., {Bensby}, T., \& {Feltzing}, S.
  2011, \aap, 530, A138

\bibitem[{{Chabrier} {et~al.}(2007){Chabrier}, {Gallardo}, \&
  {Baraffe}}]{2007chabrier}
{Chabrier}, G., {Gallardo}, J., \& {Baraffe}, I. 2007, \aap, 472, L17

\bibitem[{{Chabrier} \& {K{\"u}ker}(2006)}]{2006chabrier}
{Chabrier}, G., \& {K{\"u}ker}, M. 2006, \aap, 446, 1027

\bibitem[{{Chandrasekhar}(1961)}]{1961Chandrasekhar}
{Chandrasekhar}, S. 1961, \apj, 134, 662

\bibitem[{{Cushing} {et~al.}(2008){Cushing}, {Marley}, {Saumon}, {Kelly},
  {Vacca}, {Rayner}, {Freedman}, {Lodders}, \& {Roellig}}]{2008cushing}
{Cushing}, M.~C., {et~al.} 2008, \apj, 678, 1372

\bibitem[{{da Silva} {et~al.}(2009){da Silva}, {Torres}, {de La Reza}, {Quast},
  {Melo}, \& {Sterzik}}]{2009dasilva}
{da Silva}, L., {Torres}, C.~A.~O., {de La Reza}, R., {Quast}, G.~R., {Melo},
  C.~H.~F., \& {Sterzik}, M.~F. 2009, \aap, 508, 833

\bibitem[{{de Bruijne} \& {Eilers}(2012)}]{2012debruijne}
{de Bruijne}, J.~H.~J., \& {Eilers}, A.-C. 2012, \aap, 546, A61

\bibitem[{{de la Reza} {et~al.}(1989){de la Reza}, {Torres}, {Quast},
  {Castilho}, \& {Vieira}}]{1989delareza}
{de la Reza}, R., {Torres}, C.~A.~O., {Quast}, G., {Castilho}, B.~V., \&
  {Vieira}, G.~L. 1989, \apjl, 343, L61

\bibitem[{{de Zeeuw} {et~al.}(1999){de Zeeuw}, {Hoogerwerf}, {de Bruijne},
  {Brown}, \& {Blaauw}}]{1999dezeeuw}
{de Zeeuw}, P.~T., {Hoogerwerf}, R., {de Bruijne}, J.~H.~J., {Brown}, A.~G.~A.,
  \& {Blaauw}, A. 1999, \aj, 117, 354

\bibitem[{{Delfosse} {et~al.}(2013){Delfosse}, {Donati}, {Kouach},
  {H{\'e}brard}, {Doyon}, {Artigau}, {Bouchy}, {Boisse}, {Brun}, {Hennebelle},
  {Widemann}, {Bouvier}, {Bonfils}, {Morin}, {Moutou}, {Pepe}, {Udry}, {do
  Nascimento}, {Alencar}, {Castilho}, {Martioli}, {Wang}, {Figueira}, \&
  {Santos}}]{2013delfosse}
{Delfosse}, X., {et~al.} 2013, in SF2A-2013: Proceedings of the Annual meeting
  of the French Society of Astronomy and Astrophysics, ed. L.~{Cambresy},
  F.~{Martins}, E.~{Nuss}, \& A.~{Palacios}, 497--508

\bibitem[{{Dittmann} {et~al.}(2013){Dittmann}, {Irwin}, {Charbonneau}, \&
  {Berta-Thompson}}]{2014dittmann}
{Dittmann}, J.~A., {Irwin}, J.~M., {Charbonneau}, D., \& {Berta-Thompson},
  Z.~K. 2013, ArXiv e-prints

\bibitem[{{Donati} {et~al.}(2006){Donati}, {Catala}, {Landstreet}, \&
  {Petit}}]{2006donati}
{Donati}, J.-F., {Catala}, C., {Landstreet}, J.~D., \& {Petit}, P. 2006, in
  Astronomical Society of the Pacific Conference Series, Vol. 358, Astronomical
  Society of the Pacific Conference Series, ed. {R.~Casini \& B.~W.~Lites},
  362--+

\bibitem[{{Donati} {et~al.}(1997){Donati}, {Semel}, {Carter}, {Rees}, \&
  {Collier Cameron}}]{1997donati}
{Donati}, J.-F., {Semel}, M., {Carter}, B.~D., {Rees}, D.~E., \& {Collier
  Cameron}, A. 1997, \mnras, 291, 658

\bibitem[{{Dotter} {et~al.}(2008){Dotter}, {Chaboyer}, {Jevremovi{\'c}},
  {Kostov}, {Baron}, \& {Ferguson}}]{2008dotter}
{Dotter}, A., {Chaboyer}, B., {Jevremovi{\'c}}, D., {Kostov}, V., {Baron}, E.,
  \& {Ferguson}, J.~W. 2008, \apjs, 178, 89

\bibitem[{{Epchtein} {et~al.}(1997){Epchtein}, {de Batz}, {Capoani},
  {Chevallier}, {Copet}, {Fouqu{\'e}}, {Lacombe}, {Le Bertre}, {Pau}, {Rouan},
  {Ruphy}, {Simon}, {Tiph{\`e}ne}, {Burton}, {Bertin}, {Deul}, {Habing},
  {Borsenberger}, {Dennefeld}, {Guglielmo}, {Loup}, {Mamon}, {Ng}, {Omont},
  {Provost}, {Renault}, {Tanguy}, {Kimeswenger}, {Kienel}, {Garzon}, {Persi},
  {Ferrari-Toniolo}, {Robin}, {Paturel}, {Vauglin}, {Forveille}, {Delfosse},
  {Hron}, {Schultheis}, {Appenzeller}, {Wagner}, {Balazs}, {Holl},
  {L{\'e}pine}, {Boscolo}, {Picazzio}, {Duc}, \& {Mennessier}}]{1997epchtein}
{Epchtein}, N., {et~al.} 1997, The Messenger, 87, 27

\bibitem[{{Feiden} \& {Chaboyer}(2012)}]{2012bfeiden}
{Feiden}, G.~A., \& {Chaboyer}, B. 2012, \apj, 761, 30

\bibitem[{{Feiden} \& {Chaboyer}(2013)}]{2013feiden}
---. 2013, \apj, 779, 183

\bibitem[{{Fern{\'a}ndez} {et~al.}(2008){Fern{\'a}ndez}, {Figueras}, \&
  {Torra}}]{2008fernandez}
{Fern{\'a}ndez}, D., {Figueras}, F., \& {Torra}, J. 2008, \aap, 480, 735

\bibitem[{{Gagn{\'e}} {et~al.}(2014){Gagn{\'e}}, {Lafreni{\`e}re}, {Doyon},
  {Malo}, \& {Artigau}}]{2014gagne}
{Gagn{\'e}}, J., {Lafreni{\`e}re}, D., {Doyon}, R., {Malo}, L., \& {Artigau},
  {\'E}. 2014, \apj, 783, 121

\bibitem[{{Gastine} {et~al.}(2012){Gastine}, {Duarte}, \&
  {Wicht}}]{2012gastine}
{Gastine}, T., {Duarte}, L., \& {Wicht}, J. 2012, \aap, 546, A19

\bibitem[{{Glebocki} \& {Gnacinski}(2005)}]{2005glebocki}
{Glebocki}, R., \& {Gnacinski}, P. 2005, VizieR Online Data Catalog, 3244, 0

\bibitem[{{Gontcharov}(2006)}]{2006gontcharov}
{Gontcharov}, G.~A. 2006, Astronomy Letters, 32, 759

\bibitem[{{Gorlova} {et~al.}(2003){Gorlova}, {Meyer}, {Rieke}, \&
  {Liebert}}]{2003gorlova}
{Gorlova}, N.~I., {Meyer}, M.~R., {Rieke}, G.~H., \& {Liebert}, J. 2003, \apj,
  593, 1074

\bibitem[{{H{\o}g} {et~al.}(2000){H{\o}g}, {Fabricius}, {Makarov}, {Urban},
  {Corbin}, {Wycoff}, {Bastian}, {Schwekendiek}, \& {Wicenec}}]{2000hog}
{H{\o}g}, E., {et~al.} 2000, \aap, 355, L27

\bibitem[{{Jeffries}(2006)}]{2006jeffries}
{Jeffries}, R.~D. 2006, {Pre-Main-Sequence Lithium Depletion}, ed. S.~{Randich}
  \& L.~{Pasquini}, 163

\bibitem[{{Jenkins} {et~al.}(2009){Jenkins}, {Ramsey}, {Jones}, {Pavlenko},
  {Gallardo}, {Barnes}, \& {Pinfield}}]{2009jenkins}
{Jenkins}, J.~S., {Ramsey}, L.~W., {Jones}, H.~R.~A., {Pavlenko}, Y.,
  {Gallardo}, J., {Barnes}, J.~R., \& {Pinfield}, D.~J. 2009, \apj, 704, 975

\bibitem[{{Koen} {et~al.}(2002){Koen}, {Kilkenny}, {van Wyk}, {Cooper}, \&
  {Marang}}]{2002koen}
{Koen}, C., {Kilkenny}, D., {van Wyk}, F., {Cooper}, D., \& {Marang}, F. 2002,
  \mnras, 334, 20

\bibitem[{{Koen} {et~al.}(2010){Koen}, {Kilkenny}, {van Wyk}, \&
  {Marang}}]{2010koen}
{Koen}, C., {Kilkenny}, D., {van Wyk}, F., \& {Marang}, F. 2010, \mnras, 403,
  1949

\bibitem[{{Kraus} {et~al.}(2014){Kraus}, {Shkolnik}, {Allers}, \&
  {Liu}}]{2014kraus}
{Kraus}, A.~L., {Shkolnik}, E.~L., {Allers}, K.~N., \& {Liu}, M.~C. 2014, \aj,
  147, 146

\bibitem[{{Macdonald} \& {Mullan}(2010)}]{2010macdonald}
{Macdonald}, J., \& {Mullan}, D.~J. 2010, \apj, 723, 1599

\bibitem[{{Makarov}(2007)}]{2007makarov}
{Makarov}, V.~V. 2007, \apjs, 169, 105

\bibitem[{{Malo} {et~al.}(2013){Malo}, {Doyon}, {Lafreni{\`e}re}, {Artigau},
  {Gagn{\'e}}, {Baron}, \& {Riedel}}]{2013malo}
{Malo}, L., {Doyon}, R., {Lafreni{\`e}re}, D., {Artigau}, {\'E}., {Gagn{\'e}},
  J., {Baron}, F., \& {Riedel}, A. 2013, \apj, 762, 88

\bibitem[{{Mann} {et~al.}(2013){Mann}, {Gaidos}, \& {Ansdell}}]{2013mann}
{Mann}, A.~W., {Gaidos}, E., \& {Ansdell}, M. 2013, ArXiv e-prints

\bibitem[{{Mentuch} {et~al.}(2008){Mentuch}, {Brandeker}, {van Kerkwijk},
  {Jayawardhana}, \& {Hauschildt}}]{2008mentuch}
{Mentuch}, E., {Brandeker}, A., {van Kerkwijk}, M.~H., {Jayawardhana}, R., \&
  {Hauschildt}, P.~H. 2008, \apj, 689, 1127

\bibitem[{{Mohanty} {et~al.}(2004{\natexlab{a}}){Mohanty}, {Basri},
  {Jayawardhana}, {Allard}, {Hauschildt}, \& {Ardila}}]{2004amohanty}
{Mohanty}, S., {Basri}, G., {Jayawardhana}, R., {Allard}, F., {Hauschildt}, P.,
  \& {Ardila}, D. 2004{\natexlab{a}}, \apj, 609, 854

\bibitem[{{Mohanty} {et~al.}(2004{\natexlab{b}}){Mohanty}, {Jayawardhana}, \&
  {Basri}}]{2004bmohanty}
{Mohanty}, S., {Jayawardhana}, R., \& {Basri}, G. 2004{\natexlab{b}}, \apj,
  609, 885

\bibitem[{{Montes} {et~al.}(2001){Montes}, {L{\'o}pez-Santiago}, {G{\'a}lvez},
  {Fern{\'a}ndez-Figueroa}, {De Castro}, \& {Cornide}}]{2001montes}
{Montes}, D., {L{\'o}pez-Santiago}, J., {G{\'a}lvez}, M.~C.,
  {Fern{\'a}ndez-Figueroa}, M.~J., {De Castro}, E., \& {Cornide}, M. 2001,
  \mnras, 328, 45

\bibitem[{{Morin} {et~al.}(2008){Morin}, {Donati}, {Petit}, {Delfosse},
  {Forveille}, {Albert}, {Auri{\`e}re}, {Cabanac}, {Dintrans}, {Fares},
  {Gastine}, {Jardine}, {Ligni{\`e}res}, {Paletou}, {Ramirez Velez}, \&
  {Th{\'e}ado}}]{2008morin}
{Morin}, J., {et~al.} 2008, \mnras, 390, 567

\bibitem[{{Nidever} {et~al.}(2002){Nidever}, {Marcy}, {Butler}, {Fischer}, \&
  {Vogt}}]{2002nidever}
{Nidever}, D.~L., {Marcy}, G.~W., {Butler}, R.~P., {Fischer}, D.~A., \& {Vogt},
  S.~S. 2002, \apjs, 141, 503

\bibitem[{{Ochsenbein} {et~al.}(2000){Ochsenbein}, {Bauer}, \&
  {Marcout}}]{2000ochsenbein}
{Ochsenbein}, F., {Bauer}, P., \& {Marcout}, J. 2000, \aaps, 143, 23

\bibitem[{{Ortega} {et~al.}(2002){Ortega}, {de la Reza}, {Jilinski}, \&
  {Bazzanella}}]{2002ortega}
{Ortega}, V.~G., {de la Reza}, R., {Jilinski}, E., \& {Bazzanella}, B. 2002,
  \apjl, 575, L75

\bibitem[{{Pecaut} \& {Mamajek}(2013)}]{2013pecaut}
{Pecaut}, M.~J., \& {Mamajek}, E.~E. 2013, \apjs, 208, 9

\bibitem[{{Rajpurohit} {et~al.}(2013){Rajpurohit}, {Reyl{\'e}}, {Allard},
  {Homeier}, {Schultheis}, {Bessell}, \& {Robin}}]{2013rajpurohit}
{Rajpurohit}, A.~S., {Reyl{\'e}}, C., {Allard}, F., {Homeier}, D.,
  {Schultheis}, M., {Bessell}, M.~S., \& {Robin}, A.~C. 2013, \aap, 556, A15

\bibitem[{{Randich} {et~al.}(2001){Randich}, {Pallavicini}, {Meola},
  {Stauffer}, \& {Balachandran}}]{2001randich}
{Randich}, S., {Pallavicini}, R., {Meola}, G., {Stauffer}, J.~R., \&
  {Balachandran}, S.~C. 2001, \aap, 372, 862

\bibitem[{{Reiners}(2012)}]{2012reiners}
{Reiners}, A. 2012, Living Reviews in Solar Physics, 9, 1

\bibitem[{{Reiners} \& {Basri}(2009)}]{2009creiners}
{Reiners}, A., \& {Basri}, G. 2009, \aap, 496, 787

\bibitem[{{Reyl{\'e}} {et~al.}(2011){Reyl{\'e}}, {Rajpurohit}, {Schultheis}, \&
  {Allard}}]{2011reyle}
{Reyl{\'e}}, C., {Rajpurohit}, A.~S., {Schultheis}, M., \& {Allard}, F. 2011,
  in Astronomical Society of the Pacific Conference Series, Vol. 448, 16th
  Cambridge Workshop on Cool Stars, Stellar Systems, and the Sun, ed.
  C.~{Johns-Krull}, M.~K. {Browning}, \& A.~A. {West}, 929

\bibitem[{{Riedel} {et~al.}(2014){Riedel}, {Finch}, {Henry}, {Subasavage},
  {Jao}, {Malo}, {Rodriguez}, {White}, {Gies}, {Dieterich}, {Winters},
  {Davison}, {Nelan}, {Blunt}, {Cruz}, {Rice}, \& {Ianna}}]{2014riedel}
{Riedel}, A.~R., {et~al.} 2014, \aj, 147, 85

\bibitem[{{Rodriguez} {et~al.}(2014){Rodriguez}, {Zuckerman}, {Kastner},
  {Vican}, {Bessell}, {Faherty}, \& {Murphy}}]{2014rodriguez}
{Rodriguez}, D., {Zuckerman}, B.~M., {Kastner}, J.~H., {Vican}, L., {Bessell},
  M.~S., {Faherty}, J.~K., \& {Murphy}, S. 2014, in American Astronomical
  Society Meeting Abstracts, Vol. 223, American Astronomical Society Meeting
  Abstracts, 334.06

\bibitem[{{Rodriguez} {et~al.}(2011){Rodriguez}, {Bessell}, {Zuckerman}, \&
  {Kastner}}]{2011rodriguez}
{Rodriguez}, D.~R., {Bessell}, M.~S., {Zuckerman}, B., \& {Kastner}, J.~H.
  2011, \apj, 727, 62

\bibitem[{{Rodriguez} {et~al.}(2013){Rodriguez}, {Zuckerman}, {Kastner},
  {Bessel}, {Faherty}, \& {Murphy}}]{2013rodriguez}
{Rodriguez}, D.~R., {Zuckerman}, B., {Kastner}, J.~H., {Bessel}, M.~S.,
  {Faherty}, J.~K., \& {Murphy}, S.~J. 2013, ArXiv e-prints

\bibitem[{{Saar}(1994)}]{1994saar}
{Saar}, S.~H. 1994, in IAU Symposium, Vol. 154, Infrared Solar Physics, ed.
  D.~M. {Rabin}, J.~T. {Jefferies}, \& C.~{Lindsey}, 493

\bibitem[{{Schlieder} {et~al.}(2010){Schlieder}, {L{\'e}pine}, \&
  {Simon}}]{2010schlieder}
{Schlieder}, J.~E., {L{\'e}pine}, S., \& {Simon}, M. 2010, \aj, 140, 119

\bibitem[{{Scholz} {et~al.}(2007){Scholz}, {Coffey}, {Brandeker}, \&
  {Jayawardhana}}]{2007scholz}
{Scholz}, A., {Coffey}, J., {Brandeker}, A., \& {Jayawardhana}, R. 2007, \apj,
  662, 1254

\bibitem[{{Shkolnik} {et~al.}(2012){Shkolnik}, {Anglada-Escude}, {Liu},
  {Bowler}, {Weinberger}, {Boss}, {Reid}, \& {Tamura}}]{2012shkolnik}
{Shkolnik}, E.~L., {Anglada-Escude}, G., {Liu}, M.~C., {Bowler}, B.~P.,
  {Weinberger}, A.~J., {Boss}, A.~P., {Reid}, I.~N., \& {Tamura}, M. 2012,
  ArXiv e-prints

\bibitem[{{Shkolnik} {et~al.}(2011){Shkolnik}, {Liu}, {Reid}, {Dupuy}, \&
  {Weinberger}}]{2011shkolnik}
{Shkolnik}, E.~L., {Liu}, M.~C., {Reid}, I.~N., {Dupuy}, T., \& {Weinberger},
  A.~J. 2011, \apj, 727, 6

\bibitem[{{Siess} {et~al.}(2000){Siess}, {Dufour}, \& {Forestini}}]{2000siess}
{Siess}, L., {Dufour}, E., \& {Forestini}, M. 2000, \aap, 358, 593

\bibitem[{{Soderblom}(2010)}]{2010soderblom}
{Soderblom}, D.~R. 2010, \araa, 48, 581

\bibitem[{{Soderblom} {et~al.}(2013){Soderblom}, {Hillenbrand}, {Jeffries},
  {Mamajek}, \& {Naylor}}]{2013soderblom}
{Soderblom}, D.~R., {Hillenbrand}, L.~A., {Jeffries}, R.~D., {Mamajek}, E.~E.,
  \& {Naylor}, T. 2013, ArXiv e-prints

\bibitem[{{Song} {et~al.}(2002){Song}, {Bessell}, \& {Zuckerman}}]{2002song}
{Song}, I., {Bessell}, M.~S., \& {Zuckerman}, B. 2002, \apjl, 581, L43

\bibitem[{{Song} {et~al.}(2003){Song}, {Zuckerman}, \& {Bessell}}]{2003song}
{Song}, I., {Zuckerman}, B., \& {Bessell}, M.~S. 2003, \apj, 599, 342

\bibitem[{{Torres} {et~al.}(2000){Torres}, {da Silva}, {Quast}, {de la Reza},
  \& {Jilinski}}]{2000torres}
{Torres}, C.~A.~O., {da Silva}, L., {Quast}, G.~R., {de la Reza}, R., \&
  {Jilinski}, E. 2000, \aj, 120, 1410

\bibitem[{{Torres} {et~al.}(2006){Torres}, {Quast}, {da Silva}, {de La Reza},
  {Melo}, \& {Sterzik}}]{2006torres}
{Torres}, C.~A.~O., {Quast}, G.~R., {da Silva}, L., {de La Reza}, R., {Melo},
  C.~H.~F., \& {Sterzik}, M. 2006, \aap, 460, 695

\bibitem[{{Torres} {et~al.}(2008){Torres}, {Quast}, {Melo}, \&
  {Sterzik}}]{2008torres}
{Torres}, C.~A.~O., {Quast}, G.~R., {Melo}, C.~H.~F., \& {Sterzik}, M.~F. 2008,
  {Young Nearby Loose Associations}, ed. B.~Reipurth, 757--+

\bibitem[{{Udry} {et~al.}(2007){Udry}, {Bonfils}, {Delfosse}, {Forveille},
  {Mayor}, {Perrier}, {Bouchy}, {Lovis}, {Pepe}, {Queloz}, \&
  {Bertaux}}]{2007udry}
{Udry}, S., {et~al.} 2007, \aap, 469, L43

\bibitem[{{van Leeuwen}(2007)}]{2007vanleeuwen}
{van Leeuwen}, F., ed. 2007, Astrophysics and Space Science Library, Vol. 350,
  {Hipparcos, the New Reduction of the Raw Data}

\bibitem[{{Yee} \& {Jensen}(2010)}]{2010yee}
{Yee}, J.~C., \& {Jensen}, E.~L.~N. 2010, \apj, 711, 303

\bibitem[{{Zacharias} {et~al.}(2013){Zacharias}, {Finch}, {Girard}, {Henden},
  {Bartlett}, {Monet}, \& {Zacharias}}]{2013zacharias}
{Zacharias}, N., {Finch}, C.~T., {Girard}, T.~M., {Henden}, A., {Bartlett},
  J.~L., {Monet}, D.~G., \& {Zacharias}, M.~I. 2013, \aj, 145, 44

\bibitem[{{Zuckerman} \& {Song}(2004)}]{2004zuckerman}
{Zuckerman}, B., \& {Song}, I. 2004, \araa, 42, 685

\bibitem[{{Zuckerman} {et~al.}(2001){Zuckerman}, {Song}, {Bessell}, \&
  {Webb}}]{2001bzuckerman}
{Zuckerman}, B., {Song}, I., {Bessell}, M.~S., \& {Webb}, R.~A. 2001, \apjl,
  562, L87

\bibitem[{{Zuckerman} \& {Webb}(2000)}]{2000zuckermanwebb}
{Zuckerman}, B., \& {Webb}, R.~A. 2000, \apj, 535, 959

\end{thebibliography}

\clearpage
\LongTables
\begin{landscape}
\tabletypesize{\tiny}
\begin{deluxetable}{lrrrrrrrrrrrrrrrr}
\tablewidth{0pt}
\tablecolumns{17}
\tablecaption{Fundamental properties of YMG {\it bona fide} and candidates \label{tab:candprop}}
\tablehead{
\colhead{Name\tablenotemark{a}} & \colhead{Other} & \colhead{Spectral\tablenotemark{b}} & \colhead{vsini\tablenotemark{c}} & \colhead{RV\tablenotemark{c}} & \colhead{$P_{v}$\tablenotemark{d}} & \colhead{$d_{s}$\tablenotemark{e}} & \colhead{$d_{\pi}$\tablenotemark{f}} & \colhead{$P_{v+\pi}$\tablenotemark{d}} & \colhead{Temperature} & \colhead{Radius} & \colhead{$\log L_{bol}$} & \colhead{$\log~g$} & \colhead{[M$/$H]} & \colhead{EW Li} & \colhead{$\log~L_{\rm x}$\tablenotemark{g}} & \colhead{Parameters} \\
\colhead{(2MASS)} & \colhead{Name} & \colhead{Type} & \colhead{(km s$^{-1}$)} & \colhead{(km s$^{-1}$)} & \colhead{(\%)} & \colhead{(pc)} & \colhead{(pc)} & \colhead{(\%)} & \colhead{($K$)} & \colhead{(\Rsun)} & \colhead{(erg s$^{-1}$)} & \colhead{(dex)} & \colhead{(dex)} & \colhead{(\AA)} & \colhead{(erg s$^{-1}$)} & \colhead{Refs.}
}
\startdata
\cutinhead{BPMG bona fide}
J00275035-3233238\tablenotemark{a} 	& GJ 2006 A 	& M3.5Ve 	&     4.0 					& $    8.80\pm 0.19$ 					& $ 99.9$ 	& $  32\pm  2$ 	& $  32.3\pm  1.8$\tablenotemark{j} & $ 99.9$ & $3100\pm 18$ & $0.64\pm0.04$ & $   -1.47\pm    0.17$ & $4.50\pm0.00$ & $-0.50\pm 0.10$ 	& $ >   28.70$ 			& $   29.50\pm    0.12$ 				 &        1  \\[0.5ex]
J00275023-3233060\tablenotemark{a} 	& GJ 2006 B 	& M3.5Ve 	&     6.0 					& $    8.50\pm 0.21$ 					& $ 99.9$ 	& $  33\pm  2$ 	& $  32.3\pm  1.8$\tablenotemark{j} & $ 99.9$ & $3150\pm 18$ & $0.66\pm0.04$ & $   -1.41\pm    0.15$ & $4.50\pm0.00$ & $-0.50\pm 0.10$ 	& $ >   25.55$ 			& $   29.50\pm    0.12$ 				 &        1  \\[0.5ex]
J01112542+1526214\tablenotemark{a} 	& GJ 3076   	& M5V+M6V 	&    17.9 					& $    1.80\pm 0.20$ 					& $ 99.9$ 	& $ 21\pm 2$  	& $  21.8\pm  1.0$\tablenotemark{j}	& $ 99.9$  & $2950\pm 17$ & $0.40\pm0.02$ & $   -1.96\pm    0.17$ & $4.50\pm0.08$ & $-0.00\pm 0.22$ 	& $  592.74\pm    3.12$ & $     28.62\pm0.18$ &        1  \\[0.5ex]
J02232663+2244069 					& HIP 11152  	& M3Ve 		&     6.0\tablenotemark{l}  & $   10.40\pm 2.00$\tablenotemark{l} 	& $ 99.9$ 	& \ldots  		& $  28.7\pm  2.3$\tablenotemark{}	& $ 99.9$  & \ldots 	& \ldots 			& \ldots 				& \ldots 		& \ldots 			& $ >   21.45$ 			& $     29.58\pm0.22$ &        1  \\[0.5ex]
\ldots            					& \ldots    	& \ldots 	&   \ldots 					&    \ldots 							& \ldots  	& \ldots  		&  \ldots							& \ldots  & $3906\pm 20$ & $0.59\pm0.05$ 	& $   -1.13\pm    0.16$ & \ldots  		& \ldots  			& $    \ldots$ 			& \ldots  								 &        2  \\[0.5ex]
J02272924+3058246 					& HIP 11437 A 	& K8 		&     5.0\tablenotemark{m}  & $    6.74\pm 0.03$\tablenotemark{m} 	& $ 99.9$ 	& \ldots  		& $  40.0\pm  3.6$\tablenotemark{}	& $ 99.9$  & $4300\pm114$ & $0.82\pm0.11$ & $   -0.72\pm    0.15$ & $4.50\pm0.25$ & $-0.00\pm 0.09$ 	& $  243.88\pm    1.62$ & $     29.90\pm     0.14$ &        1  \\[0.5ex]
\ldots            					& \ldots      	& \ldots 	&   \ldots 					& \ldots 								& \ldots  	& \ldots  		& \ldots 							& \ldots  & $4359\pm 12$ & $0.84\pm0.08$ 	& $   -0.64\pm    0.18$ & \ldots  		& \ldots  			& $  220.00$\tablenotemark{u} 			& \ldots  								 &        2  \\[0.5ex]
J02412589+0559181\tablenotemark{a} 	& hip12545 AB 	& K6Ve(sb1) &    20.0\tablenotemark{i}  & $   10.00$\tablenotemark{i} 			& \ldots 	& \ldots  		& $  42.0\pm  2.7$\tablenotemark{}	& \ldots  & $4100\pm 30$ & $0.98\pm0.06$ & $   -0.62\pm    0.08$ & $4.50\pm0.16$ & $ 0.30\pm 0.00$ 	& $  445.73\pm    3.67$ & $     29.93\pm     0.14$ &        1  \\[0.5ex]
\ldots            					& \ldots      	& \ldots    &    \ldots  				& \ldots 								& \ldots  	& \ldots  		& \ldots 							& \ldots  & $4044\pm 14$ & $0.96\pm0.06$ 	& $   -0.66\pm    0.13$ & \ldots  		& \ldots  			& $  450.00$\tablenotemark{i} 			& \ldots  								 &        2  \\[0.5ex]
J04593483+0147007				 	& HIP 23200  	& M0Ve 		&    14.0\tablenotemark{i}	& $   19.82\pm    0.04$\tablenotemark{n} & $ 99.9$   & \ldots  		& $  25.9\pm  1.7$\tablenotemark{}  & $ 99.9$  & $3866\pm 18$ & $0.90\pm0.07$ 	& $   -0.79\pm    0.31$ & \ldots  		& \ldots  			& $  270.00$\tablenotemark{i} 			& \ldots  								 &        2  \\[0.5ex]
J05004714-5715255\tablenotemark{a} 	& HIP 23309  	& M0.5 kee 	&     5.8\tablenotemark{i}	& $   19.40\pm    0.30$\tablenotemark{i} & $ 99.9$   & \ldots  		& $  26.8\pm  0.8$\tablenotemark{}  & $ 99.9$  & $3884\pm 17$ & $0.93\pm0.03$ 	& $   -0.75\pm    0.06$ & \ldots  		& \ldots  			& $  360.00$\tablenotemark{i} 			& \ldots  								 &        2  \\[0.5ex]
J06131330-2742054\tablenotemark{a}	& \ldots     	& M3.5V* 	&     2.4 					& $   22.80\pm    0.20$ 				& $ 99.9$ 	& $  25\pm  6$ 	& $  29.4\pm  0.9$\tablenotemark{}  & $ 99.9$ & $3150\pm 13$ & $0.86\pm0.03$ & $   -1.18\pm    0.08$ & $4.50\pm0.00$ & $-0.50\pm 0.07$ 	& $ >   28.30$ 			& $   29.50\pm    0.07$ 				 &        1  \\[0.5ex]
J06182824-7202416 					& HIP 29964  	& K4Ve 		&    16.4\tablenotemark{i}	& $   16.20\pm    1.00$\tablenotemark{o} & $ 99.9$   & \ldots  		& $  38.6\pm  1.3$\tablenotemark{}  & $ 99.9$  & $4260\pm  9$ & $0.96\pm0.04$ 	& $   -0.56\pm    0.07$ & \ldots  		& \ldots  			& $  420.00$\tablenotemark{i} 			& \ldots  								 &        2  \\[0.5ex]
J10172689-5354265 					& TWA 22AB   	& M6Ve+M6Ve &   \ldots 					& $   13.57\pm    0.26$\tablenotemark{p} & $ 99.9$   & \ldots  		& $  17.5\pm  0.2$\tablenotemark{}  & $ 99.9$  & $2843\pm  8$ & $0.50\pm0.01$ 	& $   -1.84\pm    0.03$ & \ldots  		& \ldots  			& $  510.00$\tablenotemark{q} 			& \ldots  								 &        2  \\[0.5ex]
J20100002-2801410 					& \ldots    	& M2.5+M3.5 &    44.0 					& $   -5.90\pm    0.40$ 					& $ 99.9$ 	& $  52\pm  3$ 	& $  48.0\pm  3.1$\tablenotemark{j} & $ 99.9$ & $3100\pm 19$ & $1.22\pm0.08$ & $   -0.91\pm    0.12$ & $4.50\pm0.05$ & $-0.00\pm 0.14$ 	& $ >   45.70$ 			& $   29.60\pm    0.17$ 				 &        1  \\[0.5ex]
J20333759-2556521\tablenotemark{a}	& \ldots     	& M4.5V 	&    21.0 					& $   -8.80\pm    0.30$ 					& $ 99.9$ 	& $  40\pm  3$ 	& $  48.3\pm  3.3$\tablenotemark{j} & $ 99.9$ & $2850\pm 33$ & $0.78\pm0.06$ & $   -1.42\pm    0.20$ & $4.25\pm0.15$ & $-0.25\pm 0.28$ 	& $  504.21\pm    4.70$ & $   29.03\pm    0.35$ 				 &        1  \\[0.5ex]
J20434114-2433534\tablenotemark{a}	& \ldots     	& M3.7+M4.1 &    26.0 					& $   -6.10\pm    0.30$ 					& $ 99.9$ 	& $  44\pm  3$ 	& $  28.1\pm  3.9$\tablenotemark{k} & $ 99.9$ & $3200\pm 42$ & $0.64\pm0.10$ & $   -1.42\pm    0.47$ & $4.50\pm0.05$ & $-0.50\pm 0.24$ 	& $ >   28.25$ 			& $   29.27\pm    0.25$ 				 &        1  \\[0.5ex]
J20415111-3226073 					& HIP 102141 B  & M4Ve 		&    15.8\tablenotemark{i}	& $   -5.13\pm    0.05$\tablenotemark{n} & $ 99.9$   & \ldots  		& $  10.7\pm  0.4$\tablenotemark{}  & $ 99.9$  & $3123\pm 12$ & $0.91\pm0.04$ 	& $   -1.15\pm    0.08$ & \ldots  		& \ldots  			& $    0.00$\tablenotemark{i} 			& \ldots  								 &        2  \\[0.5ex]
J20450949-3120266\tablenotemark{a}	& HIP 102409  	& M1Ve 		&     9.3\tablenotemark{i}	& $   -4.13\pm    0.03$\tablenotemark{n} & $ 99.9$   & \ldots  		& $   9.9\pm  0.1$\tablenotemark{}  & $ 99.9$  & $3642\pm 22$ & $0.81\pm0.02$ 	& $   -0.99\pm    0.02$ & \ldots  		& \ldots  			& $   80.00$\tablenotemark{i} 			& \ldots  								 &        2  \\[0.5ex]
J22450004-3315258\tablenotemark{a} & HIP 112312 B  	& M5IVe 	&    16.8\tablenotemark{i}	& $    2.03\pm    0.04$\tablenotemark{n} & $ 99.9$   & \ldots  		& $  23.3\pm  2.0$\tablenotemark{}  & $ 99.9$  & $3031\pm 10$ & $0.56\pm0.05$ 	& $   -1.59\pm    0.17$ & \ldots  		& \ldots  			& $  450.00$\tablenotemark{i} 			& \ldots  								 &        2  \\[0.5ex]
J22445794-3315015\tablenotemark{a} & HIP 112312  	& M4IVe 	&    12.1\tablenotemark{i}	& $    3.09\pm    0.04$\tablenotemark{n} & $ 99.9$   & \ldots  		& $  23.3\pm  2.0$\tablenotemark{}  & $ 99.9$  & $3179\pm 14$ & $0.77\pm0.07$ 	& $   -1.26\pm    0.17$ & \ldots  		& \ldots  			& $    0.00$\tablenotemark{i} 			& \ldots  								 &        2  \\[0.5ex]
\cutinhead{BPMG candidate}
J00233468+2014282 					& \ldots     	& K7.5V(sb2) &    4.6 					& $   -1.60\pm 0.20$ 					& $ 65.1$ 	& $  53\pm  4$ 	& \ldots  							& \ldots  & $3900\pm 23$ & $1.18\pm0.10$ & $   -0.56\pm    0.09$ & $4.50\pm0.16$ & $-0.00\pm 0.00$ 	& $  338.65\pm    1.89$ & $   29.66\pm    0.19$\tablenotemark{h} &        1  \\[0.5ex]
J01351393-0712517 					& \ldots     	& M4V(sb2) 	&    49.6 					& $    6.30\pm 0.50$ 					& $ 88.7$ 	& $  47\pm  2$ 	& $  37.9\pm  2.4$\tablenotemark{k} & $ 75.9$ & $3100\pm 42$ & $0.78\pm0.06$ & $   -1.32\pm    0.18$ & $4.50\pm0.13$ & $-0.00\pm 0.29$ 	& $ >   46.70$ 			& $   29.32\pm    0.15$ 				 &        1  \\[0.5ex]
J01365516-0647379 					& G271-110   	& M4V+>L0 	&    10.0 					& $   12.20\pm 0.40$ 					& $ 99.9$ 	& $  21\pm  1$ 	& $  24.0\pm  0.4$\tablenotemark{k} & $ 99.9$ & $3500\pm 75$ & $0.20\pm0.01$ & $   -2.29\pm    0.17$ & $5.50\pm0.00$ & $-0.25\pm 0.22$ 	& $ >   22.85$ 			& $   28.81\pm    0.12$					 &        1  \\[0.5ex]
J03323578+2843554\tablenotemark{a}	& \ldots        & M4+M4.5 	&    21.9 					& $    9.20\pm 0.30$ 					& $ 99.9$ 	& $  55\pm  4$ 	& \ldots  							& \ldots  & $3100\pm 41$ & $0.98\pm0.08$ & $   -1.11\pm    0.18$ & $4.50\pm0.12$ & $-0.00\pm 0.30$ 	& $ >   31.35$ 			& $   29.14\pm    0.24$\tablenotemark{h} &        1  \\[0.5ex]
J04435686+3723033 					& PMI04439+3723W  & M3Ve 	&    10.6 					& $    6.40\pm 0.20$ 					& $ 96.4$ 	& $  59\pm  5$ 	& \ldots  							& \ldots  & $3700\pm243$ & $0.54\pm0.13$ & $   -1.34\pm    0.44$ & $5.00\pm0.22$ & $ 0.30\pm 0.20$ 	& $  194.12\pm    4.04$ & $   29.33\pm    0.22$\tablenotemark{h} &        1  \\[0.5ex]
J05082729-2101444 					& \ldots 		& M5V 		&    27.6 					& $   26.60\pm 0.43$ 					& $ 99.9$ 	& $  25\pm  5$ 	& \ldots  							& \ldots  & $2900\pm108$ & $0.36\pm0.09$ & $   -2.06\pm    0.99$ & $4.50\pm0.24$ & $-0.00\pm 0.30$ 	& $  481.82\pm   10.11$ & $   28.44\pm    0.37$\tablenotemark{h} &        1  \\[0.5ex]
J05241914-1601153 					& \ldots 		& M4.5+M5.0 &    50.5 					& $   17.20\pm 0.50$ 					& $ 99.9$ 	& $  20\pm  5$ 	& \ldots  							& \ldots  & $2950\pm 52$ & $0.46\pm0.11$ & $   -1.82\pm    0.98$ & $4.25\pm0.12$ & $-0.25\pm 0.25$ 	& $  464.25\pm   23.30$ & $   28.70\pm    0.33$\tablenotemark{h} &        1  \\[0.5ex]
J05335981-0221325 					& \ldots 		& M3V 		&     5.4 					& $   21.00\pm 0.20$ 					& $ 99.9$ 	& $  42\pm  5$ 	& \ldots  							& \ldots  & $3250\pm 30$ & $1.10\pm0.13$ & $   -0.94\pm    0.22$ & $4.50\pm0.05$ & $-0.50\pm 0.15$ 	& $ >   23.85$ 			& $   29.78\pm    0.17$\tablenotemark{h} &        1  \\[0.5ex]
J14252913-4113323 					& SCR1425-4113   & M2.5Ve* 	&    95.3 					& $   -1.20\pm 1.30$ 					& $ 47.9$ 	& $  60\pm  4$ 	& $  66.9\pm  4.3$\tablenotemark{}  & $ 89.2$ & $3200\pm 43$ & $1.52\pm0.12$ & $   -0.67\pm    0.11$ & $4.50\pm0.07$ & $-0.25\pm 0.24$ 	& $  684.64\pm   15.15$ &    \ldots			 				 &        1  \\[0.5ex]
J18580415-2953045\tablenotemark{a}	& TYC6872-1011-1 & M0Ve 	&    33.8\tablenotemark{i}  & $   -4.90\pm$\tablenotemark{i} 		& $ 99.9$ 	& $  76\pm  5$ 	& \ldots  							& \ldots  & $3950\pm 53$ & $0.96\pm0.06$ & $   -0.69\pm    0.09$ & $5.25\pm0.13$ & $ 0.30\pm 0.06$ 	& $  466.26\pm    5.23$ & $   30.09\pm    0.16$\tablenotemark{h} &        1  \\[0.5ex]
J19102820-2319486\tablenotemark{a}	& \ldots 		& M4V 		&    11.3 					& $   -7.20\pm 0.20$ 					& $ 99.9$ 	& $  67\pm  5$ 	& \ldots  							& \ldots  & $3350\pm 33$ & $1.04\pm0.07$ & $   -0.92\pm    0.13$ & $4.75\pm0.05$ & $-0.00\pm 0.17$ 	& $ >   23.30$ 			& $   29.88\pm    0.19$\tablenotemark{h} &        1  \\[0.5ex]
J19233820-4606316\tablenotemark{a}	& \ldots 		& M0V 		&    13.6 					& $    0.30\pm 0.26$ 					& $ 99.3$ 	& $  70\pm  4$ 	& \ldots  							& \ldots  & $4250\pm 67$ & $0.74\pm0.05$ & $   -0.78\pm    0.09$ & $5.25\pm0.14$ & $ 0.30\pm 0.15$ 	& $  425.92\pm    3.02$ & $   29.71\pm    0.23$\tablenotemark{h} &        1  \\[0.5ex]
J21100535-1919573\tablenotemark{a}	& \ldots 		& M2V 		&     9.4 					& $   -5.60\pm 0.20$ 					& $ 99.9$ 	& $  32\pm  2$ 	& \ldots  							& \ldots  & $3500\pm 60$ & $0.74\pm0.06$ & $   -1.11\pm    0.15$ & $4.75\pm0.16$ & $-0.00\pm 0.13$ 	& $ >   21.20$ 			& $   29.67\pm    0.14$\tablenotemark{h} &        1  \\[0.5ex]
J21103147-2710578\tablenotemark{a}	& \ldots 		& M4.5V 	&     9.4 					& $   -5.60\pm 0.20$ 					& $ 99.9$ 	& $  41\pm  3$ 	& \ldots  							& \ldots  & $2950\pm 24$ & $0.52\pm0.04$ & $   -1.74\pm    0.26$ & $4.50\pm0.12$ & $-0.00\pm 0.32$ 	& $  501.48\pm    9.30$ & $   28.84\pm    0.27$\tablenotemark{h} &        1  \\[0.5ex]
\cutinhead{THA bona fide}
J00240899-6211042 					& HIP 1910 AB  	& M0Ve* 	&    20.9\tablenotemark{i} 			& $    6.60\pm 0.60$\tablenotemark{i} 		& $ 99.0$  	& \ldots  		& $  53.0\pm  7.6$\tablenotemark{}  & $ 99.9$  & $3823\pm 18$ & $1.07\pm0.17$ 	& $   -0.66\pm    0.29$ & \ldots  		& \ldots  			& $  194.00$\tablenotemark{i}	& \ldots  								 &        2  \\[0.5ex]
J00251465-6130483 					& HIP 1993  	& M0Ve 		&     7.3\tablenotemark{i}  			& $    6.40\pm 0.10$\tablenotemark{i}   	& $ 99.0$  	& \ldots  		& $  45.8\pm  5.1$\tablenotemark{}  & $ 99.9$  & $4015\pm 14$ & $0.78\pm0.10$ 	& $   -0.85\pm    0.22$ & \ldots  		& \ldots  			& $   40.00$\tablenotemark{i}	& \ldots  								 &        2  \\[0.5ex]
J00345120-6154583 					& HIP 2729  	& K5Ve 		&   122.8\tablenotemark{i}			& $   -1.00\pm 2.00$\tablenotemark{q}   	& $ 99.0$  	& \ldots  		& $  43.9\pm  1.9$\tablenotemark{}  & $ 99.9$  & $4376\pm 10$ & $1.19\pm0.06$ 	& $   -0.33\pm    0.09$ & \ldots  		& \ldots  			& $  360.00$\tablenotemark{i}	& \ldots  								 &        2  \\[0.5ex]
J00452814-5137339 					& HIP 3556  	& M3V 		&     5.0\tablenotemark{r}			& $   -1.60\pm19.99$\tablenotemark{s}		& $ 11.0$  	& \ldots  		& $  40.4\pm  4.3$\tablenotemark{}  & $ 90.0$  & $3576\pm 23$ & $0.82\pm0.10$ 	& $   -1.00\pm    0.21$ & \ldots  		& \ldots  			& $   55.00$\tablenotemark{t}	& \ldots  								 &        2  \\[0.5ex]
J21443012-6058389 					& HIP 107345  	& M1V 		&     8.2\tablenotemark{i}			& $    2.30\pm 0.50$\tablenotemark{i}   	& $ 99.0$  	& \ldots  		& $  43.6\pm  4.9$\tablenotemark{}  & $ 99.9$  & $3843\pm 18$ & $0.72\pm0.09$ 	& $   -0.99\pm    0.22$ & \ldots  		& \ldots  			& $   55.00$ 			& \ldots  								 &        2  \\[0.5ex]
\cutinhead{THA candidate}
J01220441-3337036\tablenotemark{a}	& \ldots 		& K7Ve 		&     5.0 					& $    5.00\pm 0.20$ 				 	& $ 99.9$ 	& $  39\pm  2$ 	& \ldots  							& \ldots  & $4400\pm 61$ & $0.59\pm0.04$ & $   -0.96\pm    0.10$ & $5.25\pm0.13$ & $-0.00\pm 0.00$ 	& $ >   19.50$ 			& $   29.62\pm    0.13$\tablenotemark{h} &        1  \\[0.5ex]
J02001277-0840516\tablenotemark{a}	& \ldots 		& M2.5V 	&    15.0 					& $    4.80\pm 0.20$ 				 	& $ 91.3$ 	& $  38\pm  2$ 	& \ldots  							& \ldots  & $3250\pm 32$ & $0.92\pm0.05$ & $   -1.07\pm    0.12$ & $4.50\pm0.12$ & $-0.25\pm 0.16$ 	& $ >   25.05$ 			& $   29.42\pm    0.15$\tablenotemark{h} &        1  \\[0.5ex]
J02155892-0929121\tablenotemark{a}	& \ldots 		& M2.5+M5+M8 &   16.7 					& $    8.30\pm 0.30$ 				 	& $ 99.9$ 	& $  44\pm  3$ 	& \ldots  							& \ldots  & $3350\pm 32$ & $1.00\pm0.08$ & $   -0.94\pm    0.15$ & $5.00\pm0.00$ & $-0.25\pm 0.18$ 	& $   24.40\pm    4.85$ & $   29.67\pm    0.14$\tablenotemark{h} &        1  \\[0.5ex]
J04365738-1613065\tablenotemark{a}	& \ldots 		& M3.5V 	&    50.7 					& $   15.60\pm 0.50$ 				 	& $ 99.3$ 	& $  47\pm  4$ 	& \ldots  							& \ldots  & $3350\pm 84$ & $0.72\pm0.07$ & $   -1.21\pm    0.17$ & $5.00\pm0.18$ & $-0.00\pm 0.27$ 	& $ >   48.20$ 			& $   29.90\pm    0.15$\tablenotemark{h} &        1  \\[0.5ex]
\cutinhead{COL bona fide}
J03413724+5513068\tablenotemark{a}	& HIP~17248  	& M0.5V 	&     5.0 					& $   -3.20$ 						 	& $ 99.9$ 	&  \ldots		&  $ 35.2\pm 2.7$\tablenotemark{} 	& $ 99.9$  & $4050\pm 36$ & $0.60\pm0.05$ & $   -1.06\pm    0.17$ & $5.25\pm0.23$ & $ 0.30\pm 0.00$ 	& $ >   19.35$ 			& \ldots &        1  \\[0.5ex]
\cutinhead{COL candidate}
J01373940+1835332 					& TYC 1208-468-1  & K3V+K5V &    16.6 					& $    0.90\pm 0.30$ 				 	& $ 95.8$ 	& $  63\pm  3$  & \ldots 							& \ldots  & $4350\pm193$ & $1.50\pm0.23$ & $   -0.15\pm    0.05$ & $5.00\pm0.54$ & $-0.00\pm 0.15$ 	& $  440.80\pm    3.74$ & \ldots &        1  \\[0.5ex]
J02335984-1811525\tablenotemark{a}	& \ldots 		& M3.0+M3.5 &    16.4 					& $   12.40\pm 0.30$ 				 	& $ 99.9$ 	& $  77\pm  5$ 	& \ldots  							& \ldots  & $3350\pm 43$ & $0.84\pm0.07$ & $   -1.09\pm    0.16$ & $5.00\pm0.07$ & $-0.00\pm 0.21$ 	& $ >   27.55$ 			& $   29.80\pm    0.19$\tablenotemark{h} &        1  \\[0.5ex]
J04071148-2918342\tablenotemark{a}	& \ldots 		& K7.5+M1.0 &    19.6 					& $   21.20\pm 0.70$ 				 	& $ 99.9$ 	& $  71\pm  5$ 	& \ldots  							& \ldots  & $3800\pm 67$ & $1.04\pm0.08$ & $   -0.71\pm    0.09$ & $5.00\pm0.11$ & $ 0.30\pm 0.15$ 	& $  286.83\pm    4.22$ & $   29.82\pm    0.16$\tablenotemark{h} &        1  \\[0.5ex]
J05100427-2340407 					& \ldots 		& M3+M3.5 	&     8.2 					& $   24.20\pm 0.30$ 				 	& $ 99.9$ 	& $  49\pm  6$ 	& \ldots  							& \ldots  & $3300\pm 33$ & $0.88\pm0.11$ & $   -1.11\pm    0.26$ & $4.75\pm0.08$ & $-0.00\pm 0.24$ 	& $ >   22.95$ 			& $   29.72\pm    0.19$\tablenotemark{h} &        1  \\[0.5ex]
J05142878-1514546 					& \ldots 		& M3.5V(vb) &     7.2					& $   21.50\pm 0.20$ 				 	& $ 98.9$ 	& $  58\pm  8$ 	& \ldots  							& \ldots  & $3100\pm 13$ & $0.48\pm0.08$ & $   -1.72\pm    0.54$ & $4.50\pm0.00$ & $-0.25\pm 0.14$ 	& $ >   23.25$ 			& $   29.48\pm    0.27$\tablenotemark{h} &        1  \\[0.5ex]
J05241317-2104427 					& \ldots 		& M4V 		&     6.5 					& $   24.70\pm 0.20$ 				 	& $ 99.3$ 	& $  50\pm  6$ 	& \ldots  							& \ldots  & $3150\pm 27$ & $0.54\pm0.06$ & $   -1.59\pm    0.35$ & $4.50\pm0.00$ & $-0.50\pm 0.16$ 	& $ >   23.55$ 			& $   29.12\pm    0.29$\tablenotemark{h} &        1  \\[0.5ex]
J23314492-0244395 					& \ldots 		& M4.5V 	&     5.5 					& $   -5.30\pm 0.20$ 				 	& $ 56.3$ 	& $  45\pm  2$ 	& \ldots  							& \ldots  & $2900\pm 19$ & $0.82\pm0.04$ & $   -1.37\pm    0.13$ & $4.25\pm0.10$ & $-0.50\pm 0.17$ 	& $ >   28.00$ 			& $   29.46\pm    0.15$\tablenotemark{h} &        1  \\[0.5ex]
\cutinhead{ARG candidate}
J09445422-1220544\tablenotemark{a}	& NLTT 22503 	& M5V 		&    36.0 					& $   13.50\pm 0.40$ 				 	& $ 99.9$ 	& $  12\pm  1$ 	& $ 13.9\pm 0.3$\tablenotemark{k} 	& $ 99.9$ & $2850\pm 25$ & $0.40\pm0.02$ & $   -2.03\pm    0.15$ & $4.50\pm0.18$ & $-0.00\pm 0.31$ 	& $ >   41.25$ 			& $   28.63\pm    0.11$\tablenotemark{h} &        1  \\[0.5ex]
J18450097-1409053 					& \ldots 		& M5V(vb) 	&    13.7 					& $  -23.50\pm 0.20$ 				 	& $ 99.9$ 	& $  16\pm  2$ 	& \ldots  							& \ldots  & $3000\pm 23$ & $0.46\pm0.06$ & $   -1.84\pm    0.48$ & $4.50\pm0.05$ & $-0.25\pm 0.20$ 	& $ >   26.30$ 			& $   28.78\pm    0.24$\tablenotemark{h} &        1  \\[0.5ex]
J19224278-0515536 					& \ldots 		& K5V 		&     9.4 					& $  -26.30\pm 0.20$ 				 	& $ 99.9$ 	& $  45\pm  6$ 	& \ldots  							& \ldots  & $3200\pm 35$ & $0.56\pm0.07$ & $   -1.53\pm    0.37$ & $4.75\pm0.11$ & $-0.50\pm 0.22$ 	& $ >   23.65$ 			& $   29.30\pm    0.26$\tablenotemark{h} &        1  \\[0.5ex]
J20163382-0711456\tablenotemark{a}	& \ldots 		& M0V+M2V 	&     5.7 					& $  -23.00\pm 0.20$ 				 	& $ 99.9$ 	& $  33\pm  3$ 	& \ldots  							& \ldots  & $3850\pm 68$ & $0.58\pm0.06$ & $   -1.14\pm    0.19$ & $5.25\pm0.20$ & $-0.00\pm 0.16$ 	& $ >   21.15$ 			& $   28.78\pm    0.25$\tablenotemark{h} &        1  \\[0.5ex]
\cutinhead{ABDMG bona fide}
J03472333-0158195 					& HIP~17695 	& M2.5V kee &    18.0 					& $   16.00$ 				 	& $ 99.9$ 	& \ldots  		& $ 16.1\pm 0.7$\tablenotemark{} 	& $ 99.9$  & $3350\pm 40$ & $0.48\pm0.03$ & $   -1.57\pm    0.15$ & $5.00\pm0.00$ & $-0.25\pm 0.21$ 	& $   23.10\pm    1.69$ & \ldots &        1  \\[0.5ex]
J22232904+3227334 					& HIP~110526 AB	& M3V* 		&    16.0 					& $  -20.60$ 				 	& $ 68.0$ 	& \ldots  		& $ 15.5\pm 1.6$\tablenotemark{} 	& $ 99.9$  & $3350\pm 43$ & $0.68\pm0.07$ & $   -1.28\pm    0.26$ & $5.00\pm0.04$ & $-0.00\pm 0.24$ 	& $   24.95\pm    4.50$ & \ldots &        1  \\[0.5ex]
J23060482+6355339 					& HIP~114066 	& M1V 		&     8.0 					& $  -23.70$ 				 	& $ 99.9$	& \ldots  		& $ 24.5\pm 1.0$\tablenotemark{} 	& $ 99.9$  & $3950\pm 77$ & $0.56\pm0.03$ & $   -1.20\pm    0.10$ & $5.25\pm0.14$ & $ 0.30\pm 0.14$ 	& $   33.10\pm    2.18$ & \ldots &        1  \\[0.5ex]
\cutinhead{ABDMG candidate}
J01123504+1703557\tablenotemark{a}	& GuPsc 		& M3V	 	&    22.5 					& $   -1.60\pm 0.30$ 				 	& $ 99.9$ 	& $  48\pm  2$ 	& \ldots  							& \ldots  & $3250\pm 32$ & $0.54\pm0.04$ & $   -1.53\pm    0.19$ & $4.75\pm0.07$ & $-0.25\pm 0.19$ 	& $ >   30.10$ 			& $   29.13\pm    0.23$\tablenotemark{h} &        1  \\[0.5ex]
J04571728-0621564\tablenotemark{a}	& \ldots 		& M0.5V 	&    11.1 					& $   23.50\pm 0.30$ 				 	& $ 99.9$ 	& $  49\pm  3$ 	& \ldots  							& \ldots  & $3950\pm 66$ & $0.52\pm0.04$ & $   -1.22\pm    0.15$ & $5.25\pm0.16$ & $ 0.30\pm 0.12$ 	& $ >   19.05$ 			& $   29.13\pm    0.25$\tablenotemark{h} &        1  \\[0.5ex]
J10285555+0050275\tablenotemark{a}	& HIP~51317 	& M2V 		&     1.1 					& $    8.30\pm 0.30$ 				 	& $ 99.9$ 	& \ldots  		& $ 7.06\pm 0.02$\tablenotemark{} 	& $ 99.9$  & $3500\pm 44$ & $0.44\pm0.01$ & $   -1.60\pm    0.06$ & $5.00\pm0.13$ & $-0.25\pm 0.18$ 	& $ >   24.25$ 			& \ldots &        1  \\[0.5ex]
J12383713-2703348\tablenotemark{a}	& \ldots 		& M2.5V 	&     4.4 					& $    9.60\pm 0.20$ 				 	& $ 99.9$ 	& $  25\pm  1$ 	& \ldots  							& \ldots  & $3400\pm 58$ & $0.48\pm0.04$ & $   -1.55\pm    0.18$ & $5.00\pm0.14$ & $-0.25\pm 0.22$ 	& $ >   23.55$ 			& $   28.81\pm    0.19$\tablenotemark{h} &        1  \\[0.5ex]
J20465795-0259320\tablenotemark{a}	& \ldots 		& M0V 		&     9.5 					& $  -14.10\pm 0.40$ 				 	& $ 99.9$ 	& $  46\pm  2$ 	& \ldots  							& \ldots  & $4050\pm 45$ & $0.56\pm0.03$ & $   -1.14\pm    0.10$ & $5.25\pm0.20$ & $ 0.30\pm 0.00$ 	& $   26.22\pm    3.75$ & $   29.19\pm    0.24$\tablenotemark{h} &        1  \\[0.5ex]
J23320018-3917368\tablenotemark{a}	& \ldots 		& M3V 		&     5.6 					& $   10.90\pm 0.20$ 				 	& $ 99.9$ 	& $  23\pm  1$ 	& \ldots  							& \ldots  & $3350\pm 57$ & $0.40\pm0.02$ & $   -1.76\pm    0.15$ & $5.00\pm0.08$ & $-0.00\pm 0.23$ 	& $ >   23.75$ 			& $   29.18\pm    0.18$\tablenotemark{h} &        1  \\[0.5ex]
J23513366+3127229\tablenotemark{a}	& \ldots 		& M2V+L0 	&    12.9 					& $  -13.60\pm 0.30$ 				 	& $ 99.9$ 	& $  42\pm  2$ 	& \ldots  							& \ldots  & $3300\pm 43$ & $0.54\pm0.03$ & $   -1.51\pm    0.16$ & $4.25\pm0.12$ & $-0.25\pm 0.18$ 	& $ >   22.10$ 			& $   29.15\pm    0.19$\tablenotemark{h} &        1  \\[0.5ex]
\enddata
\tablenotetext{a}{Stars used for further analysis.}
\tablenotetext{b}{Spectral type with asterisk is for the whole unresolved system.}
\tablenotetext{c}{Measured $\vsini$ and Radial velocity (RV) using ESPaDOnS spectrum (see paper II), unless stated otherwise.}
\tablenotetext{d}{Membership probability including radial velocity information ($P_v$), or membership probability including radial velocity and parallax information (P$_{v+\pi}$).}
\tablenotetext{e}{Statistical distance derived by our analysis (see Section 5 of paper I).}
\tablenotetext{f}{Parallax measurement from \citet{2007vanleeuwen}, unless stated otherwise.}
\tablenotetext{g}{X-ray luminosity using the parallax measurement, unless stated otherwise.}
\tablenotetext{h}{X-ray luminosity using the statistical distance.}
\tablecomments{(i) \citet{2006torres}; (j) \citet{2014riedel}; (k) \citet{2012shkolnik}; (l) \citet{2010schlieder}; (m) \citet{2003song}; (n) \citet{2012bailey}; (o) \citet{2001montes}; (p) \citet{2011shkolnik}; (q) \citet{2008fernandez}; 
 ( r) \citet{2006bobylev}; (s) \citet{2007scholz}; (t) \citet{2008mentuch}; (u) \citet{2009dasilva}.}
\tablerefs{(1) This work; (2) \citet{2013pecaut}.}
\end{deluxetable}
\clearpage
\end{landscape}
    
\clearpage
\LongTables
\begin{landscape}
\tabletypesize{\tiny}
\begin{deluxetable}{lrrrrrrrrrrrrrrrr}
\tablewidth{0pt}
\tablecolumns{17}
\tablecaption{Fundamental properties of Field stars \label{tab:fieldprop}}
\tablehead{
\colhead{Name\tablenotemark{a}} & \colhead{Other} & \colhead{Spectral} & \colhead{vsini\tablenotemark{b}} & \colhead{RV\tablenotemark{b}} & \colhead{$P_{v}$\tablenotemark{c}} & \colhead{$d_{s}$\tablenotemark{d}} & \colhead{$d_{\pi}$\tablenotemark{e}} & \colhead{$P_{v+\pi}$\tablenotemark{c}} & \colhead{Temperature} & \colhead{Radius} & \colhead{$\log L_{bol}$} & \colhead{$\log~g$} & \colhead{[M$/$H]} & \colhead{EW Li} & \colhead{$\log~L_{\rm x}$\tablenotemark{f}} & \colhead{Parameters} \\
\colhead{(2MASS)} & \colhead{Name} & \colhead{Type} & \colhead{(km s$^{-1}$)} & \colhead{(km s$^{-1}$)} & \colhead{(\%)} & \colhead{(pc)} & \colhead{(pc)} & \colhead{(\%)} & \colhead{($K$)} & \colhead{(\Rsun)} & \colhead{(erg s$^{-1}$)} & \colhead{(dex)} & \colhead{(dex)} & \colhead{(\AA)} & \colhead{(erg s$^{-1}$)} & \colhead{Refs.}
}
\startdata
\cutinhead{Field bona fide}
J00182256+4401222 	& GJ 15A  	& M1.5 V & $    2.90$\tablenotemark{i} 	& $   11.81\pm    0.10$\tablenotemark{n} & $\ldots$ & $\ldots$ & $    3.57\pm    0.01$\tablenotemark{k} & $\ldots$ & $3563\pm 11$ & $0.39\pm0.00$ & $   -1.66\pm   -0.02$ &\ldots 			& $-0.36$ 			& \ldots 		& $\ldots$ &        2  \\[0.5ex]
J05312734-0340356 	& GJ 205  	& M1.5 V & $    1.00$\tablenotemark{i} 	& $    8.66\pm    0.10$\tablenotemark{n} & $\ldots$ & $\ldots$ & $    5.66\pm    0.04$\tablenotemark{l} & $\ldots$ & $3801\pm  9$ & $0.57\pm0.00$ & $   -1.21\pm   -0.02$ &\ldots 			& $ 0.35$ 			& \ldots 		& $\ldots$ &        2  \\[0.5ex]
\ldots 				& \ldots 	& \ldots &     \ldots 					& \ldots					 			 & $\ldots$ & \ldots   & $\ldots$ 								& \ldots   & $3700\pm 75$ & $0.58\pm0.03$ & $   -1.23\pm    0.04$ & $5.50\pm0.12$ 	& $-0.00\pm 0.21$ 	& $ >   22.95$ 	& $\ldots$ &        1  \\[0.5ex]
J09142298+5241125 	& GJ 338A  	& M0.0 V & $    2.90$\tablenotemark{i} 	& $   11.14\pm    0.10$\tablenotemark{n} & $\ldots$ & $\ldots$ & $    5.81\pm    0.21$\tablenotemark{}  & $\ldots$ & $3907\pm 35$ & $0.58\pm0.01$ & $   -1.16\pm   -0.04$ &\ldots 			& $-0.18$ 			& \ldots 		& $\ldots$ &        2  \\[0.5ex]
J09142485+5241118 	& GJ 338B  	& K7.0 V & $    2.80$\tablenotemark{i} 	& $   12.49\pm    0.10$\tablenotemark{n} & $\ldots$ & $\ldots$ & $    5.81\pm    0.21$\tablenotemark{}  & $\ldots$ & $3867\pm 37$ & $0.57\pm0.01$ & $   -1.19\pm   -0.04$ &\ldots 			& $-0.15$ 			& \ldots 		& $\ldots$ &        2  \\[0.5ex]
J10112218+4927153 	& GJ 380  	& K7.0 V & $    2.70$\tablenotemark{j} 	& $  -25.73\pm    0.10$\tablenotemark{n} & $\ldots$ & $\ldots$ & $    4.87\pm    0.01$\tablenotemark{l} & $\ldots$ & $4081\pm 15$ & $0.64\pm0.00$ & $   -0.99\pm   -0.01$ &\ldots 			& $-0.16$ 			& \ldots 		& $\ldots$ &        2  \\[0.5ex]
J11032023+3558117 	& GJ 411  	& M2.0 V & $    2.90$\tablenotemark{i} 	& $  -84.69\pm    0.10$\tablenotemark{n} & $\ldots$ & $\ldots$ & $    2.55\pm    0.00$\tablenotemark{l} & $\ldots$ & $3465\pm 17$ & $0.39\pm0.00$ & $   -1.70\pm   -0.01$ &\ldots 			& $-0.41$ 			& \ldots 		& $\ldots$ &        2  \\[0.5ex]
\ldots 				& \ldots 	& \ldots &     \ldots 					& \ldots					 			 & $\ldots$ & \ldots   & $\ldots$ 								& \ldots   & $3500\pm 33$ & $0.42\pm0.01$ & $   -1.63\pm    0.03$ & $5.00\pm0.08$ 	& $-0.50\pm 0.16$ 	& $ >   24.15$ 	& $\ldots$ &        1  \\[0.5ex]
J11052903+4331357 	& GJ 412A  	& M1.0 V & $    3.00$\tablenotemark{i} 	& $   68.89\pm    0.10$\tablenotemark{n} & $\ldots$ & $\ldots$ & $    4.85\pm    0.02$\tablenotemark{m} & $\ldots$ & $3497\pm 39$ & $0.40\pm0.01$ & $   -1.67\pm   -0.02$ &\ldots 			& $-0.40$ 			& \ldots 		& $\ldots$ &        2  \\[0.5ex]
J13454354+1453317 	& GJ 526  	& M1.5 V & $    1.40$\tablenotemark{i} 	& $   15.81\pm    0.10$\tablenotemark{n} & $\ldots$ & $\ldots$ & $    5.39\pm    0.03$\tablenotemark{m} & $\ldots$ & $3618\pm 31$ & $0.48\pm0.01$ & $   -1.44\pm   -0.02$ &\ldots 			& $-0.30$ 			& \ldots 		& $\ldots$ &        2  \\[0.5ex]
\ldots 				& \ldots 	& \ldots &     2.00 					& \ldots 					 			 & $\ldots$ & \ldots   & $\ldots$ 								& \ldots   & $3750\pm 44$ & $0.44\pm0.01$ & $   -1.49\pm    0.04$ & $5.00\pm0.19$ 	& $-0.00\pm 0.16$ 	& $ >   24.25$ 	& $\ldots$ &        1  \\[0.5ex]
J17362594+6820220 	& GJ 687  	& M3.0 V & $    2.80$\tablenotemark{i} 	& $  -28.78\pm    0.10$\tablenotemark{n} & $\ldots$ & $\ldots$ & $    4.53\pm    0.02$\tablenotemark{m} & $\ldots$ & $3413\pm 28$ & $0.42\pm0.01$ & $   -1.67\pm   -0.02$ &\ldots 			& $-0.09$ 			& \ldots 		& $\ldots$ &        2  \\[0.5ex]
\ldots 				& \ldots 	& \ldots &     \ldots					& \ldots 					 			 & $\ldots$ & \ldots   & $\ldots$ 								& \ldots   & $3300\pm 28$ & $0.44\pm0.01$ & $   -1.69\pm    0.04$ & $4.75\pm0.10$ 	& $-0.25\pm 0.13$ 	& $ >   27.90$ 	& $\ldots$ &        1  \\[0.5ex]
J17574849+0441405 	& GJ 699  	& M4.0 V & $    2.80$\tablenotemark{i} 	& $ -110.51\pm    0.10$\tablenotemark{n} & $\ldots$ & $\ldots$ & $    1.82\pm    0.01$\tablenotemark{m} & $\ldots$ & $3224\pm 10$ & $0.19\pm0.00$ & $   -2.47\pm   -0.02$ &\ldots 			& $-0.39$ 			& \ldots 		& $\ldots$ &        2  \\[0.5ex]
J18052735+0229585 	& GJ 702B  	& K5 Ve & $    2.70$\tablenotemark{j} 	& $   -7.70\pm    0.20$\tablenotemark{o} & $\ldots$ & $\ldots$ & $    5.08\pm    0.02$\tablenotemark{l} & $\ldots$ & $4393\pm149$ & $0.67\pm0.01$ & $   -0.82\pm   -0.11$ &\ldots 			& $ 0.03$ 			& \ldots 		& $\ldots$ &        2  \\[0.5ex]
J18424666+5937499 	& GJ 725A  	& M3.0 V & $    5.00$\tablenotemark{i} 	& $   -0.83\pm    0.10$\tablenotemark{n} & $\ldots$ & $\ldots$ & $    3.57\pm    0.03$\tablenotemark{}  & $\ldots$ & $3407\pm 15$ & $0.36\pm0.00$ & $   -1.82\pm   -0.02$ &\ldots 			& $-0.49$ 			& \ldots 		& $\ldots$ &        2  \\[0.5ex]
J18424688+5937374 	& GJ 725B  	& M3.5 V & $    7.00$\tablenotemark{i} 	& $    1.10\pm    0.10$\tablenotemark{n} & $\ldots$ & $\ldots$ & $    3.57\pm    0.03$\tablenotemark{}  & $\ldots$ & $3104\pm 28$ & $0.32\pm0.01$ & $   -2.06\pm   -0.03$ &\ldots 			& $-0.36$ 			& \ldots 		& $\ldots$ &        2  \\[0.5ex]
J20531977+6209156 	& GJ 809  	& M0.5 & $    2.80$\tablenotemark{i} 	& $  -17.16\pm    0.10$\tablenotemark{n} & $\ldots$ & $\ldots$ & $    7.05\pm    0.03$\tablenotemark{}  & $\ldots$ & $3692\pm 22$ & $0.55\pm0.01$ & $   -1.30\pm   -0.02$ &\ldots 			& $-0.21$ 			& \ldots 		& $\ldots$ &        2  \\[0.5ex]
J22563497+1633130 	& GJ 880  	& M1.5 V & $    2.80$\tablenotemark{i} 	& $  -27.32\pm    0.10$\tablenotemark{n} & $\ldots$ & $\ldots$ & $    6.85\pm    0.05$\tablenotemark{}  & $\ldots$ & $3713\pm 11$ & $0.55\pm0.00$ & $   -1.29\pm   -0.02$ &\ldots 			& $ 0.06$ 			& \ldots 		& $\ldots$ &        2  \\[0.5ex]
\ldots 				& \ldots 	& \ldots &     \ldots 					& \ldots				 				 & $\ldots$ & \ldots   & $\ldots$ 								& \ldots   & $3700\pm 66$ & $0.52\pm0.02$ & $   -1.35\pm    0.03$ & $5.25\pm0.12$ 	& $-0.00\pm 0.15$ 	& $ >   21.05$ 	& $\ldots$ &        1  \\[0.5ex]
\cutinhead{Field candidate}
J06022455-1634494 	& \ldots 	& M0V 		&     9.10 					& $   -8.20\pm 0.20$ 					 & $ 99.90$ & $ 41\pm 9$ & \ldots  								& \ldots  	& $3600\pm 37$ & $0.70\pm0.18$ & $   -1.13\pm    0.55$ & $5.00\pm0.13$ & $-0.00\pm 0.12$ 	& $27.93\pm3.89$ & $   29.21\pm    0.31$\tablenotemark{g} &        1  \\[0.5ex]
J09361593+3731456 	& HIP~47133 & M2+M2 	&     1.90 					& $    0.00\pm 0.50$ 					 & $ 98.59$ & $ 30\pm 6$ & $  33.7\pm  2.6$					 	& $ 96.79$ 	& $3900\pm 22$ & $0.72\pm0.05$ & $   -0.97\pm    0.12$ & $5.25\pm0.17$ & $-0.50\pm 0.00$ 	& $ >   19.25$ 	& $   28.88\pm    0.23$ &        1  \\[0.5ex]
J12194808+5246450 	& HIP~60121	& K7V 		&     3.80 					& $   -4.20\pm 0.20$ 					 & $ 99.90$ & $ 32\pm 6$ & $  28.0\pm  1.7$					 	& $ 99.90$ 	& $3950\pm170$ & $0.60\pm0.12$ & $   -1.11\pm    0.34$ & $5.00\pm0.35$ & $-0.00\pm 0.18$ 	& $ >   18.70$ 	& $   28.61\pm    0.25$ &        1  \\[0.5ex]
J15594729+4403595 	& \ldots 	& M1V 		&    54.90 					& $  -15.80\pm 0.50$ 					 & $ 99.90$ & $ 20\pm 5$ & \ldots  								& \ldots  	& $3600\pm 86$ & $0.42\pm0.11$ & $   -1.60\pm    0.81$ & $5.00\pm0.14$ & $-0.00\pm 0.32$ 	& $ >   47.10$ 	& $   28.85\pm    0.34$\tablenotemark{g} &        1  \\[0.5ex]
J18495543-0134087 	& \ldots 	& M2.5V(sb1) &    34.60 				& $  116.60\pm 0.50$ 					 & $ 99.90$ & $ 23\pm 6$ & \ldots  								& \ldots  	& $3400\pm 74$ & $0.32\pm0.10$ & $   -1.94\pm    1.19$ & $5.25\pm0.14$ & $-0.25\pm 0.21$ 	& $ >   34.20$ 	& $   28.77\pm    0.37$\tablenotemark{g} &        1  \\[0.5ex]
J19420065-2104051 	& \ldots 	& M3.5V(sb2) &     2.70 				& $  -21.90\pm 0.30$ 					 & $ 99.90$ & $  8\pm 1$ & \ldots  								& \ldots  	& $3200\pm 59$ & $0.16\pm0.05$ & $   -2.62\pm    1.44$ & $5.00\pm0.16$ & $-0.00\pm 0.31$ 	& $ >   26.00$ 	& $   27.44\pm    0.44$\tablenotemark{g} &        1  \\[0.5ex]
J20531465-0221218 	& NLTT50066 & M3+M4 	&    10.00 					& $  -39.90\pm1.1$\tablenotemark{q} 	 & $ 99.90$ & $ 25\pm 6$ & $  37.9\pm  5.7$\tablenotemark{q} 	& $ 99.90$ 	& $3550\pm111$ & $0.46\pm0.09$ & $   -1.51\pm    0.49$ & $5.50\pm0.18$ & $-0.00\pm 0.24$ 	& $ >   23.65$ 	& $   28.88\pm    0.30$ &        1  \\[0.5ex]
J21073678-1304581 	& \ldots 	& M3V 		&    52.20 					& $   -2.30\pm 0.50$ 					 & $ 69.30$ & $ 17\pm 4$ & \ldots  								& \ldots  	& $3350\pm 60$ & $0.34\pm0.10$ & $   -1.88\pm    1.05$ & $5.00\pm0.12$ & $-0.25\pm 0.27$ 	& $ >   49.75$ 	& $   28.68\pm    0.33$\tablenotemark{g} &        1  \\[0.5ex]
J23172807+1936469 	& GJ~4326 	& M3.5+M4.5 &     6.70 					& $    4.40\pm 0.20$ 					 & $ 99.90$ & $  7\pm 1$ & $  9.1\pm0.2$\tablenotemark{p} 		& \ldots  	& $3250\pm 68$ & $0.26\pm0.01$ & $   -2.12\pm    0.13$ & $4.75\pm0.18$ & $-0.25\pm 0.27$ 	& $ >   26.65$ 	& $   28.13\pm    0.30$\tablenotemark{g} &        1  \\[0.5ex]
\enddata
\tablenotetext{a}{Stars used for further analysis.}
\tablenotetext{b}{Measured $\vsini$ and Radial velocity (RV) using ESPaDOnS spectrum (see paper II), unless stated otherwise.}
\tablenotetext{c}{Membership probability including radial velocity information ($P_v$), or membership probability including radial velocity and parallax information (P$_{v+\pi}$).}
\tablenotetext{d}{Statistical distance derived by our analysis (see Section 5 of paper I).}
\tablenotetext{e}{Parallax measurement from \citet{2007vanleeuwen}, unless stated otherwise.}
\tablenotetext{f}{X-ray luminosity using the parallax measurement, unless stated otherwise.}
\tablenotetext{g}{X-ray luminosity using the statistical distance.}
\tablecomments{(i) \citet{2009jenkins}; (j) \citet{2005glebocki}; (k) \citet{2001montes}; (l) \citet{2011casagrande}; (m) \citet{2012debruijne}; (n) \citet{2002nidever}; (o) \citet{2006gontcharov}; (p) \citet{2014dittmann}; (q) \citet{2012shkolnik}}
\tablerefs{(1) This work; (2) \citet{2012boyajian}.}
\end{deluxetable}
\clearpage
\end{landscape}

\end{document}